\documentclass[superscriptaddress,nofootinbib,aps,pra,twocolumn]{revtex4-2}

\usepackage{amsmath}
\usepackage{amssymb}
\usepackage{amsfonts}
\usepackage{bm} % bold math
\usepackage{bbm}
\usepackage{braket}
\usepackage{color}
\usepackage{comment}
\usepackage{dcolumn} % align table columns on decimal point
\usepackage{dsfont}
\usepackage{enumerate}
\usepackage{epsfig}
\usepackage{esint}
\usepackage[T1]{fontenc}
\usepackage{framed}
\usepackage{gensymb}
\usepackage{graphicx} % include figure files
\usepackage[colorlinks,linkcolor=blue,citecolor=blue,urlcolor=blue,hyperindex,driverfallback=dvipdfm]{hyperref}
\usepackage{indentfirst}
\usepackage{lmodern}
\usepackage{mathrsfs}
\usepackage{mathtools}
\usepackage{multirow}
\usepackage{psfrag}
\usepackage{pst-all}
\usepackage{soul}
\usepackage{xcolor}
\usepackage{xspace}
\usepackage{orcidlink}

\newcommand{\abs}[1] {\mathopen{}\left|#1\right|\mathclose{}}

\newcommand{\ccpar}[1] {\mathopen{}\left(#1\right)\mathclose{}}
\newcommand{\sqpar}[1] {\mathopen{}\left[#1\right]\mathclose{}}
\newcommand{\clpar}[1] {\mathopen{}\left\{#1\right\}\mathclose{}}

\def\ii{{\rm i}}  \def\ee{{\rm e}}
  \def\kB{{k_{\rm B}}}
  \def\Imm{{\rm Im}}
\newcommand{\pd}[2] {\mathopen{}\frac{\partial#1}{\partial#2}\mathclose{}}

\def\rb{{\bf r}}  \def\Rb{{\bf R}}    \def\vb{{\bf v}}
    
\def\xx{\hat{\bf x}}  \def\yy{\hat{\bf y}}  
    
\def\kb{{\bf k}}    
  \def\Qb{{\bf Q}}
\def\Eb{{\bf E}}        % Maxwell
\def\Jb{{\bf J}}
\def\pb{{\bf p}}
\def\vF{v_{\rm F}}    \def\EF{{E_{\rm F}}}  % solid-state, Fermi energy, etc.
  \def\kkb{\hat{\bf k}}  \def\fk{f_{\bf k}}
\def\eps{\epsilon}  \def\vep{\varepsilon}
\def\ww{\omega}
\def\Dm{\mathcal{D}}  \def\Em{\mathcal{E}}
\def\Hm{\mathcal{H}}  \def\Mm{\mathcal{M}}
\def\epsa{\epsilon_{\rm a}}  \def\epsb{\epsilon_{\rm b}}  \def\epseff{\epsilon_{\rm eff}}
  \def\phiext{\phi^{\rm ext}}  % induced and external scalar potential
  \def\Phiext{\Phi^{\rm ext}}
\def\rhoind{\rho^{\rm ind}}

\begin{document}

\title{Nonreciprocal plasmons in one-dimensional carbon nanostructures}

\author{A.~Rodr\'{\i}guez~Echarri\,\orcidlink{0000-0003-4634-985X}}
\affiliation{Max-Born-Institut, 12489 Berlin, Germany}
\affiliation{Center for Nanophotonics, NWO Institute AMOLF, 1098 XG Amsterdam, The Netherlands}

\author{F.~Javier~Garc\'{\i}a~de~Abajo\,\orcidlink{0000-0002-4970-4565}}
\affiliation{ICFO-Institut de Ciencies Fotoniques, The Barcelona Institute of Science and Technology, 08860 Castelldefels (Barcelona), Spain}
\affiliation{ICREA-Instituci\'o Catalana de Recerca i Estudis Avan\c{c}ats, Passeig Llu\'{\i}s Companys 23, 08010 Barcelona, Spain}

\author{Joel~D.~Cox\,\orcidlink{0000-0002-5954-6038}}
\email[Corresponding author: ]{cox@mci.sdu.dk}
\affiliation{POLIMA---Center for Polariton-driven Light--Matter Interactions, University of Southern Denmark, Campusvej 55, DK-5230 Odense M, Denmark}
\affiliation{Danish Institute for Advanced Study, University of Southern Denmark, Campusvej 55, DK-5230 Odense M, Denmark}

\begin{abstract}
    The directional control of light in miniaturized plasmonic waveguides holds appealing possibilities for emerging nanophotonic technologies, but is hindered by the intrinsic reciprocal optical response of conventional plasmonic materials. While the ability of graphene to sustain large electrical currents shows promise for nonreciprocal plasmonics, studies have been limited to extended samples characterized by linear electrical dispersion. Here, we theoretically explore quantum finite-size and nonlocal effects in the nonreciprocal response of mesoscale plasmonic waveguides comprised of drift-biased graphene nanoribbons (GNRs) and carbon nanotubes (CNTs). Using atomistic simulation methods based on tight-binding electronic states and self-consistent mean-field optical response, we reveal that a moderate electrical bias can significantly break reciprocity for propagation of guided plasmon modes in GNRs and CNTs exhibiting electronic band gaps. The excitation by a nearby point dipole emitter and subsequent propagation of guided plasmon modes can thus be actively controlled by the applied current, which can further be leveraged to mediate nonlocal interactions of multiple emitters. Our results establish graphene nanostructures as a promising atomically thin platform for nonreciprocal nanophotonics.
\end{abstract}

\date{\today}
\maketitle
%\tableofcontents

%%%%%%%%%%%%%%%%%%%%%%%%%%%%%%%%%%%%%%%%%%%%%%%%%%%%%%%%%%%%%%%%%%%%%%%%%%%%%%%%
\section{Introduction}

Plasmon polaritons at metal-dielectric interfaces have been widely explored to control the flow of electromagnetic energy on nanometer length scales \cite{hohenester2020nano}. Ordinarily, the directionality of light propagation in plasmonic materials is governed by Lorentz reciprocity, which ensures a symmetric response when the light source and detector are exchanged \cite{potton2004reciprocity}. However, nonreciprocal responses in plasmonics, where electromagnetic wave propagation differs depending on direction, enable important functionalities such as optical isolation~\cite{chin2013nonreciprocal,davoyan2014electrically}, unidirectional waveguiding~\cite{yu2008oneway,bliokh2018electric,li2024unidirectional}, and control of radiative heat flow~\cite{hassani2022drifting,liu2022thermal}.

Reciprocity can be broken by applying static magnetic fields~\cite{jalas2013what}, where the enhanced light-matter interactions associated with plasmon resonances can boost the otherwise intrinsically low magneto-optical response of noble metals~\cite{chin2013nonreciprocal}. In an alternative approach, plasmon-driven nonlinear optical phenomena can trigger nonreciprocal effects~\cite{boroviks2023demonstration}, although strategies for nonreciprocal photonics based on nonlinear optics present inherent limitations associated with unwanted backward-propagating noise~\cite{shi2015limitations,mahmoud2015all}. These constraints underscore the need for compact nanophotonic platforms that exhibit robust nonreciprocal behavior without relying on cumbersome magnetic biasing or power-intensive nonlinear optical processes~\cite{papaj2020plasmonic,hassani2022drifting,monticone2025nonlocality}.

Electrically doped graphene has emerged as a promising two-dimensional (2D) material platform for plasmonics, supporting plasmon resonances with high spatial dispersion that can be actively tuned by injecting charge carriers~\cite{gonccalves2016introduction}. Beyond enabling active modulation of plasmon resonances through electrostatic gating, the ability of graphene to sustain ultrahigh current densities~\cite{murali2009breakdown,liao2010high,son2017graphene,barajas2024electrically} can be harnessed for nonreciprocal plasmonics, circumventing the use of large applied magnetic fields~\cite{sabbaghi2015drift,vanduppen2016current,wenger2018current}. Experimental studies of graphene subjected to applied direct currents have demonstrated that drifting charge carriers in the carbon monolayer can appreciably break the reciprocity of plasmon propagation along the direction of current flow \cite{dong2021fizeau,zhao2021efficient,dong2025current}.

Nonreciprocal plasmon polaritons in drift-biased graphene have been widely studied in a semiclassical electrodynamic framework, adopting nonlocal 2D conductivity models obtained by populating the conical electronic band structure of graphene with a skewed Fermi-Dirac distribution \cite{sabbaghi2015drift,vanduppen2016current,wenger2018current,blevins2024plasmon} or by introducing a Doppler-shifted frequency in the local conductivity~\cite{morgado2017negative,morgado2018drift,morgado2020nonlocal}. In analogy to the Doppler shift, Fizeau shifts in the plasmon wavelength of drift-biased graphene have been experimentally measured using scanning near-field optical microscopy \cite{dong2021fizeau,zhao2021efficient}. In the presence of an applied direct current, graphene plasmons launched by localized light sources, such as point dipoles and energetic free electrons, are predicted to more efficiently transport optical energy in a preferential direction \cite{correas2019nonreciprocal,prudencio2021asymmetric,morgado2022directional}.

The pattering of graphene on nanometer length scales offers further possibilities to tailor plasmon resonances and engineer nanoscale light-matter interactions. In particular, one-dimensional (1D) carbon nanostructures, such as graphene nanoribbons (GNRs) and carbon nanotubes (CNTs), have been recognized as ideal candidates for plasmonic waveguiding that can be tuned passively by selecting the structure size or actively by modifying charge carrier doping levels \cite{christensen2012graphene,huidobro2012superradiance,martinmoreno2015ultraefficient,soto2015plasmons,devega2016plasmons,calajo2023nonlinear}. In the mesoscopic regime, quantum confinement effects in the electronic band structure of these systems can significantly impact not only their transport properties~\cite{baringhaus2014exceptional,laird2015quantum,wang2021graphene,yao2023n}, but also the associated plasmonic response \cite{thongrattanasiri2012quantum,cox2016quantum,devega2020strong}. These effects have yet to be explored in the dispersion of plasmon polariton modes supported by 1D carbon nanostructures with $\lesssim10$\,nm lateral size, with plasmons in GNRs and CNTs exhibiting high sensitivity to edge termination (i.e., armchair or zigzag) and structural chirality, respectively.

\begin{figure*}
    \centering
    \includegraphics[width=1\textwidth]{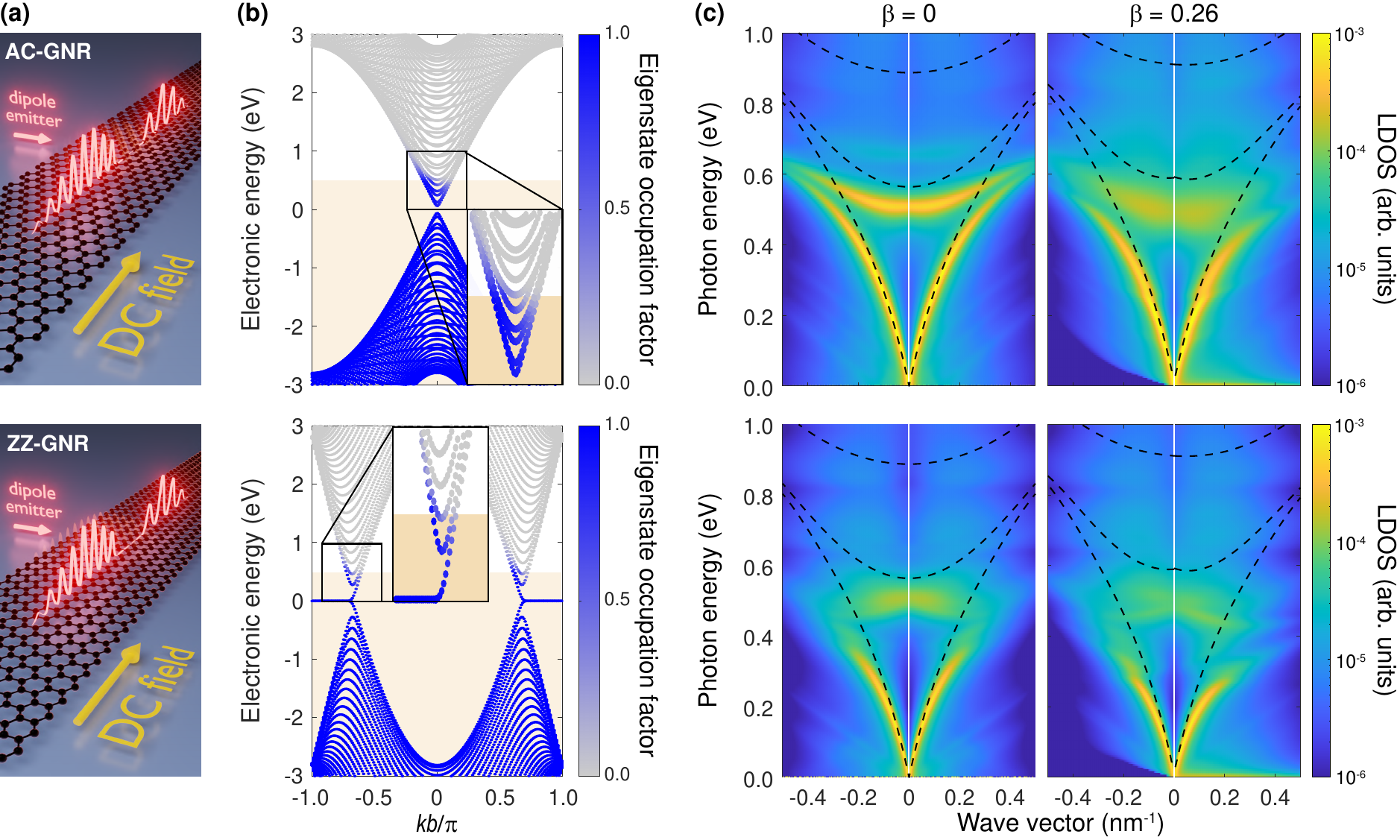}
    \caption{\textbf{Nonreciprocal plasmons in drift-biased graphene nanoribbons.}
    (a) Guided plasmon polaritons are launched by a point dipole emitter and propagate with different dispersion relative to the applied direct current (DC) in graphene nanoribbons (GNRs) with either armchair (AC, upper panel) or zigzag (ZZ, lower panel) edge terminations. (b) The one-dimensional electronic band structures of AC (upper panel) and ZZ (lower panel) GNRs of $W\approx10$\,nm width are plotted as functions of the Bloch wave vector $k$ in the direction of lattice translational symmetry, normalized in each case to the respective GNR unit cell period $b$. In the absence of an applied DC field ($E^{\rm DC}=0$), the bands are occupied up to the Fermi energy $\EF=0.5$\,eV (shaded orange region), while the occupation factors of GNRs for a static field of amplitude $E^{\rm DC}= 5.2\times 10^6$~V/m are indicated by the colors of the bands (see colorbar). (c) Wave-vector-resolved local photonic density of states (LDOS) for AC (upper panels) and ZZ (lower panels) GNRs obtained from atomistic simulations (contour plots) in the absence ($E^{\rm DC}=0$) and presence ($E^{\rm DC}=5.2\times 10^6$~V/m) of an applied DC field, with corresponding plasmon dispersions predicted by the semiclassical electrodynamic model for the indicated drift velocities $v=\beta\vF$ (dashed curves).
    }
\label{fig:GNRs}
\end{figure*}

\begin{figure}
    \centering
    \includegraphics[width=0.95\columnwidth]{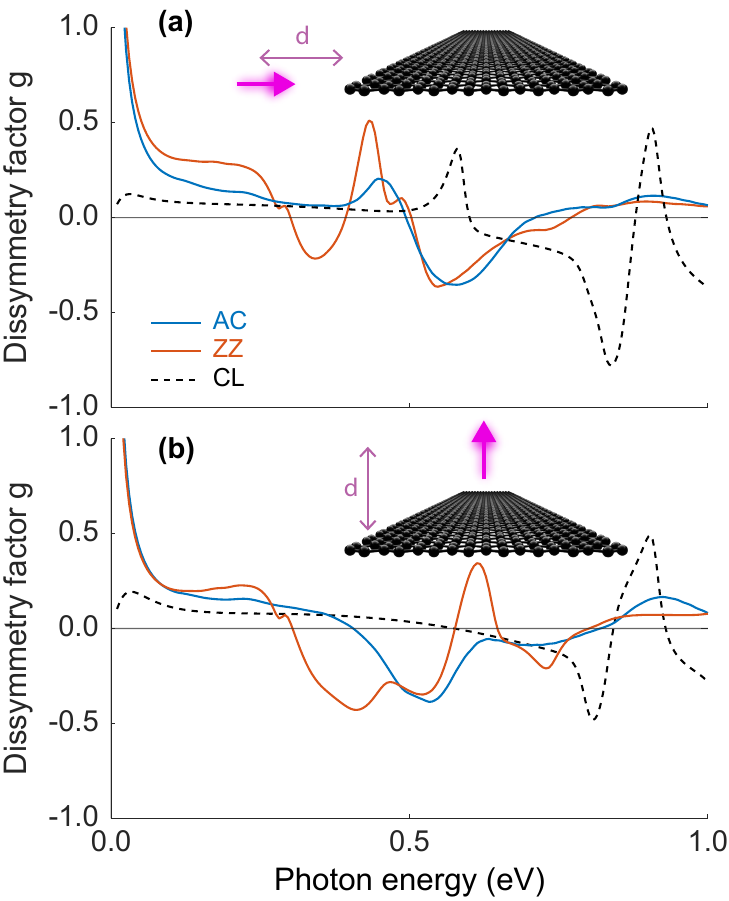}
    \caption{\textbf{Dissymmetry of guided plasmon propagation in drift-biased graphene nanoribbons.} The dissymmetry factor of Eq.~\eqref{eq:g_factor} is calculated through atomistic simulations (solid curves) of the armchair (AC, blue curves) and zigzag (ZZ, red curves) edge-terminated graphene nanoribbons (GNRs) of $W \approx 10$\,nm width considered in Fig.~\ref{fig:GNRs} for (a) a dipole oriented parallel to and placed in the GNR plane, at a distance $d=5$\,nm from the GNR edge, and (b) a dipole oriented normally to the ribbon plane and centered $d=5$\,nm above the GNR. The dissymmetry factors predicted by semiclassical simulations of the same configurations are also shown (CL, dashed curves).
    }
\label{fig:g_GNRs}
\end{figure}

Here, we explore nonreciprocal plasmon polaritons in drift-biased GNRs and CNTs. Our analysis is based on an atomistic tight-binding description of carbon nanostructures that captures salient features in the electronic band structure due to edge termination in GNRs and structural chirality in CNTs. Nonlocal effects in the optical response of 1D carbon nanostructures to a point dipole excitation source are incorporated in the framework of the random-phase approximation (RPA), employed here to resolve the dispersion of waveguided plasmons propagating along or against a drifting equilibrium electron distribution. Atomistic results predict a much greater dependence on drift current compared to estimates based on an electrodynamic description of analogous 1D carbon nanostructures, adopting a bulk 2D conductivity obtained from a semiclassical description of drift-biased graphene. In particular, topological effects in the electronic structure of zigzag GNRs and CNTs are shown to strongly impact the nonreciprocal response. Our results indicate that a strong imbalance of plasmon emission in opposite directions along these 1D carbon nanostructures, which can be controlled by the direction and magnitude of the applied currents, establishes these systems as a viable platform for active nonreciprocal nanophotonics.

%%%%%%%%%%%%%%%%%%%%%%%%%%%%%%%%%%%%%%%%%%%%%%%%%%%%%%%%%%%%%%%%%%%%%%%%%%%%%%%%
\section{Results and discussion}

In what follows, we simulate the local density of optical states (LDOS) using atomistic and semiclassical descriptions of GNRs and CNTs under an applied drift bias. The LDOS at $\rb_0=(x_0,y_0,z_0)$ is characterized by the self-interaction of a point dipole with moment $\pb$ that oscillates with frequency $\omega$, so we have $\text{LDOS} \propto\Imm\clpar{\pb^*\cdot\Eb^{\rm ind}(\rb_0,\ww)}$, where $\Eb^{\rm ind}$ denotes the electric field produced by the nanostructure in response to the dipole at $\rb_0$. To study the nonreciprocal response of 1D systems characterized by translational invariance in the $y$-direction, we decompose the induced field as $\Eb^{\rm ind}(\rb_0,\ww)=(2\pi)^{-1}\int{\rm d}q\vec{\Em}_q(x_0,z_0,\ww)\ee^{\ii q y_0}$ (i.e., in terms of spatial Fourier components $\vec{\Em}_q$), and define the wave-vector-resolved LDOS function \cite{hohenester2020nano}
\begin{equation} \label{eq:LDOS_q}
    \Dm_q(x_0,z_0,\ww) = \frac{1}{2 \pi^2 \omega |\pb|^2} \Imm\clpar{\pb^*\cdot\vec{\Em}_q(x_0,z_0,\ww)} .
\end{equation}
The directional emission of a dipole near a GNR or CNT can then be quantified by defining a dissymmetry factor as
\begin{equation} \label{eq:g_factor}
    g = 2\frac{\sum_\pm\int_0^{\pm\infty}{\rm d}q\Dm_q}{\int_{-\infty}^\infty{\rm d}q\Dm_q}
\end{equation}
[i.e., the relative difference of contributions to the LDOS from waves propagating with positive ($q>0$) and negative ($q<0$) wave vectors].

%%%%%%%%%%%%%%%%%%%%%%%%%%%%%%%%%%%%%%%%%%%%%%%%%%%%%%%%%%%%%%%%%%%%%%%%%%%%%%%%
\subsection{Graphene nanoribbons}

We consider guided plasmon polaritons launched by a nearby point dipole source in drift-biased graphene nanoribbons (GNRs) with finite width $W$ along $\xx$, lying in the $z=0$ plane, and having exclusively armchair (AC) or zigzag (ZZ) edge terminations, as illustrated schematically in the upper and lower panels of Fig.~\ref{fig:GNRs}a, respectively. In this scenario, guided plasmons that propagate along (against) the drift current parallel to $\yy$ are Doppler shifted to longer (shorter) wavelengths. Accordingly, the rate at which the dipole excites waveguided plasmons will exhibit a directional dependence. To explore this phenomenon, we calculate the LDOS in the framework of the RPA from the tight-binding (TB) electronic Bloch states of the considered AC and ZZ GNRs (see Methods). In Fig.~\ref{fig:GNRs}b, we plot the energies $\hbar\vep_{j,k}$ of electronic bands indexed by $j$ as a function of Bloch wave vector $k$ in the direction of translational lattice symmetry for GNRs of width $W\approx10$\,nm (grey dots). Following the prescription we introduce in Methods, the effect of drift currents induced by an in-plane electric field $E^{\rm DC}$ polarized along $\yy$ (the direction of translational invariance) is contained in the electronic occupation factors $f_{j,k}$ satisfying the steady-state limit of the Boltzmann transport equation,
\begin{equation} \label{eq:f_E_DC}
    \ccpar{1-\frac{e\tau}{\hbar}E^{\rm DC}\pd{}{k}}f_{j,k} = f_{j,k}^{(0)} ,
\end{equation}
where $\tau$ is a phenomenological relaxation time introduced here to account for inelastic scattering (see Methods), and
\begin{equation} \label{eq:f_0}
    f_{j,k}^{(0)}=\left[\ee^{(\hbar\vep_{j,k}-\mu)/\kB T}+1\right]^{-1}
\end{equation}
are the occupation factors in the absence of any drift current, specified by the Fermi-Dirac distribution for a chemical potential $\mu$ and an electron temperature $T$. In practice, Eq.~\eqref{eq:f_E_DC} is solved for a discrete set of wave vectors $k$ within the first Brillouin zone of each band by discretizing the differential operator and imposing periodic boundary conditions. For GNRs doped with additional charge carriers corresponding to $\mu=0.5$\,eV at room temperature $T=300$\,K, such that states with energies $\hbar\vep_{j,k}\lesssim\mu$ (indicated by the orange shaded regions in Fig.~\ref{fig:GNRs}b) are occupied when $E^{\rm DC}=0$, the drifting electron distribution induced by an applied field $E^{\rm DC}=5.2\times10^6$\,V/m is indicated by the color of the electronic bands in Fig.~\ref{fig:GNRs}b (see vertical colorbar). Significant asymmetry in the occupation factors appears around the Dirac points in drift-biased GNRs, resulting in a distinguishable optical response for fields with positive or negative in-plane optical momenta. In the Supporting Information (SI), we present calculations of the electronic band structures and occupation factors of the considered $W\approx10$\,nm AC (Fig.~S1) and ZZ (Fig.~S2) GNRs under different configurations of charge carrier doping and applied field strength.

To explore the effects of drift current on the GNR plasmon dispersion, we simulate the LDOS in Fig.~\ref{fig:GNRs}c for the $W\approx10$\,nm AC (upper panel) and ZZ (lower panel) GNRs, obtained in each case for a dipole polarized along $\xx$ and positioned 5\,nm directly above the ribbon edge. Prominent features associated with guided plasmon polaritons appearing in the LDOS show that their dispersion is independent of propagation direction in the absence of a drift current (left panels). In general, the ZZ GNR exhibits a weaker plasmonic response due to the presence of zero-energy edge states in the electronic spectrum~\cite{thongrattanasiri2012quantum}, while these modes also decay more rapidly with increasing wave vector due to Landau damping. In the presence of a finite DC field, the drift current not only induces an asymmetry in the plasmon dispersion, but also damps the guided plasmon modes. Specifically, in both AC and ZZ GNRs, the low-energy monopole mode is diminished at high optical wave vectors and the dipolar GNR modes are generally quenched. Incidentally, the plasmon dispersion relations obtained from atomistic simulations are in reasonable agreement with the predictions based on a classical electrostatic model (see Methods), which are indicated in Fig.~\ref{fig:GNRs}c by the dashed curves. To compare atomistic and classical models of drift-biased GNRs, we estimate the drift velocity $v=\beta\vF$ (expressed here in terms of the Fermi velocity $\vF\approx c/300$ in extended graphene) from $v=\mu^{\rm DC}E^{\rm DC}$, where $\mu^{\rm DC}$ is the electron mobility, which determines the inelastic scattering rate $\tau=\mu^{\rm DC}\EF/(e\vF^2)$ \cite{garciadeabajo2014graphene}. Unless otherwise specified, we assume a conservative value for the electron mobility $\mu^{\rm DC}=500$\,cm$^2$/(V\,s), while additional details on the interdependence of the drift velocity, electron mobility, and charge carrier doping level entering the classical model are presented in the SI (see Fig.~S3).

The nonreciprocal response of the drift-biased GNRs is quantified according to Eq.~\eqref{eq:g_factor} by integrating the LDOS data shown in Fig.~\ref{fig:GNRs}c over either exclusively positive ($q>0$) or negative ($q<0$) in-plane optical wave vectors. The resulting dissymmetry factors shown in Fig.~\ref{fig:g_GNRs}a indicate a distinguishable response for AC (blue curve) and ZZ (red curve) GNRs predicted by atomistic simulations, reflecting the distinct plasmon dispersion relations shown in the LDOS plots of Fig.~\ref{fig:GNRs}c, while the qualitative behavior of the emission into forward- and backward-propagating modes is captured by the classical simulations (dashed black curve) at higher energies. In Fig.~\ref{fig:g_GNRs}b we plot the dissymmetry factors $g$ obtained when the dipole is instead oriented normally to the ribbon plane and centered 5\,nm above the GNR (see inset), which, as shown in the SI, modifies the wave-vector-dependent coupling to waveguided plasmon modes predicted in atomistic (Fig.~S4) and classical (Fig.~S5) simulations, and thus also the spectral dependence of the corresponding dissymmetry factors. Qualitatively similar dissymmetry factors $g$ are obtained at higher energies when the dipole is aligned with the ribbon and centered above it (Figs.~S6), a configuration for which symmetry prohibits coupling of the lowest-order plasmon mode to the dipole.

\begin{figure*}
    \centering
    \includegraphics[width=1\textwidth]{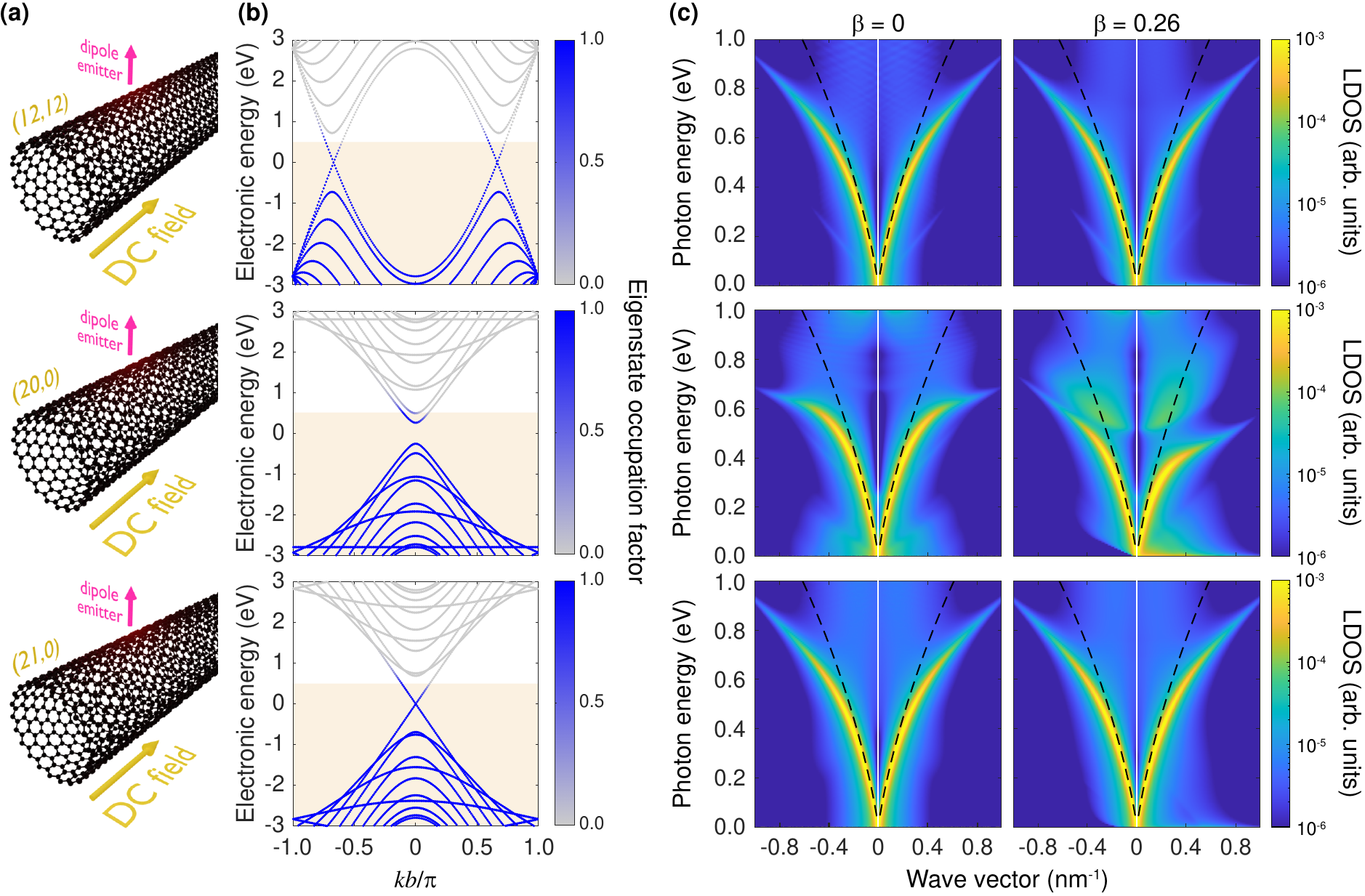}
    \caption{\textbf{Nonreciprocal plasmons in drift-biased carbon nanotubes.} For CNTs with diameter $D\sim1.6$\,nm characterized by chiralities (12,12), (20, 0), and (21,0) in the upper, middle, and lower rows, respectively, we present (a) schematic illustrations of the nanotubes interacting with a point dipole emitter. (b) Electronic band structure as a function of the Bloch wave vector $k$ in the direction of translational invariance (grey curves), normalized in each case to the respective CNT unit cell period $b$, with the shaded orange region indicating fully occupied bands up to $\EF=0.5$\,eV doping in the absence of an applied DC field ($E^{\rm DC}=0$), while the occupation factors for a static field of amplitude $E^{\rm DC}=5.2\times 10^6$~V/m are indicated by the colors of the bands (see colorbar). (c) Wave-vector-resolved local photonic density of states (LDOS) obtained from atomistic simulations (contour plots) and presented along with the corresponding plasmon dispersion predicted from a classical electrodynamic model (dashed curves) are shown in the absence of a DC field (left column) and a DC field with amplitude $E^{\rm DC}=5.2\times 10^6$~V/m (right column). In all LDOS calculations, the dipole is considered to be oriented normally to and a distance $d=5$\,nm away from the tube axes.
    }
\label{fig:CNT}
\end{figure*}

\begin{figure}
    \centering
    \includegraphics[width=0.95\columnwidth]{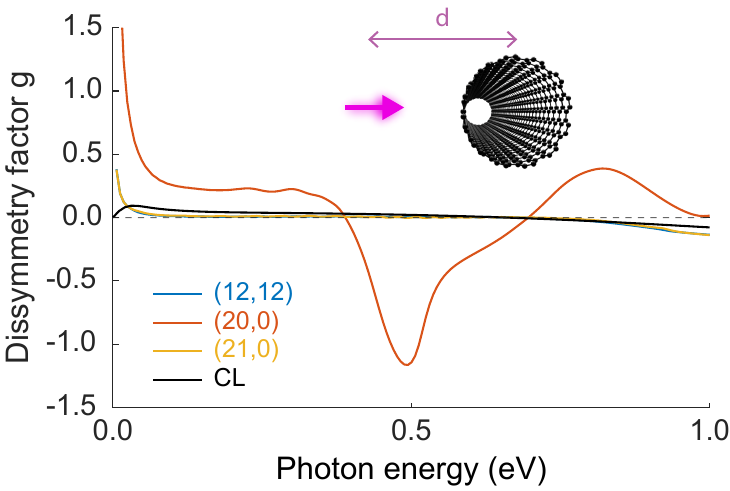}
    \caption{\textbf{Dissymmetry factors for plasmons in drift-biased CNTs.} The dissymmetry factors for the (12,12), (20,0), and (21,0) electrically doped and drift-biased CNTs considered in Fig.~\ref{fig:CNT} are plotted as blue, red, and yellow curves, respectively, under excitation by a dipole placed at a distance $d=5$~nm away from the center and oriented along the radial direction (see inset). The corresponding dissymmetry predicted from a classical model of a 1.6~nm diameter CNT (dashed curve) is shown for comparison.
    }
\label{fig:fig_dism_CNT}
\end{figure}

%%%%%%%%%%%%%%%%%%%%%%%%%%%%%%%%%%%%%%%%%%%%%%%%%%%%%%%%%%%%%%%%%%%%%%%%%%%%%%%%
\subsection{Carbon nanotubes}

The structural arrangement of carbon atoms in CNTs is characterized by indices ($n$,$m$) that define the tube type, with electronic structure varying from metallic to insulating. In Fig.~\ref{fig:CNT}, we simulate the plasmonic response of metallic (12,12), moderately semiconducting (20,0), and small-gap semiconducting (21,0) CNTs with diameters $D=(0.246\,{\rm nm}/\pi)\sqrt{n^2+m^2+mn}$, which are all approximately $D \sim 1.6$\,nm. Fig.~\ref{fig:CNT}a illustrates the considered CNTs, which support guided polaritons that can be launched by a nearby point dipole source and may exhibit directional-dependent dispersion according to the applied DC field. In Fig.~\ref{fig:CNT}b, we present the electronic bands (curves) of the considered CNTs, uniformly populated up to a Fermi energy $\EF=0.5$\,eV in the absence of a drift current (shaded orange region), and occupied according to Eq.~\eqref{eq:f_E_DC} (see colorbar) in the presence of a static field $E^{\rm DC}=5.2\times 10^6$~V/m. In the SI, we present the occupation factors of these CNTs under different carrier doping levels (see Fig.~S7). The wave-vector-resolved LDOS for a point dipole placed 5\,nm away from the CNT centers is shown in Figs.~\ref{fig:CNT}c for unbiased (left) and biased (right) CNTs, respectively, superimposing the classical dispersion (dashed curves) on the results obtained from atomistic simulations (contours). 

The wave-vector-resolved LDOS in Fig.~\ref{fig:CNT}c indicates that the nonreciprocal plasmonic response of drift-biased CNTs is negligible for doped metallic and semimetallic CNTs, while the semiconducting drift-biased CNT exhibits a large nonreciprocal response, presumably due to the drastic difference in the available virtual electron-hole-pair excitations that configure the plasmon when the plasmon wave vectors are directed to the right or left sides (i.e., along or against the region of depletion in the electronic population of the conduction band). These conclusions are supported by the corresponding dissymmetry factors shown in Fig.~\ref{fig:fig_dism_CNT}a, while a more significant dissymmetry appears in the response of all considered CNTs at higher doping levels (i.e., when the noted difference in virtual excitations for right or left propagation are exacerbated). 

\begin{figure*}
    \centering
    \includegraphics[width=1\textwidth]{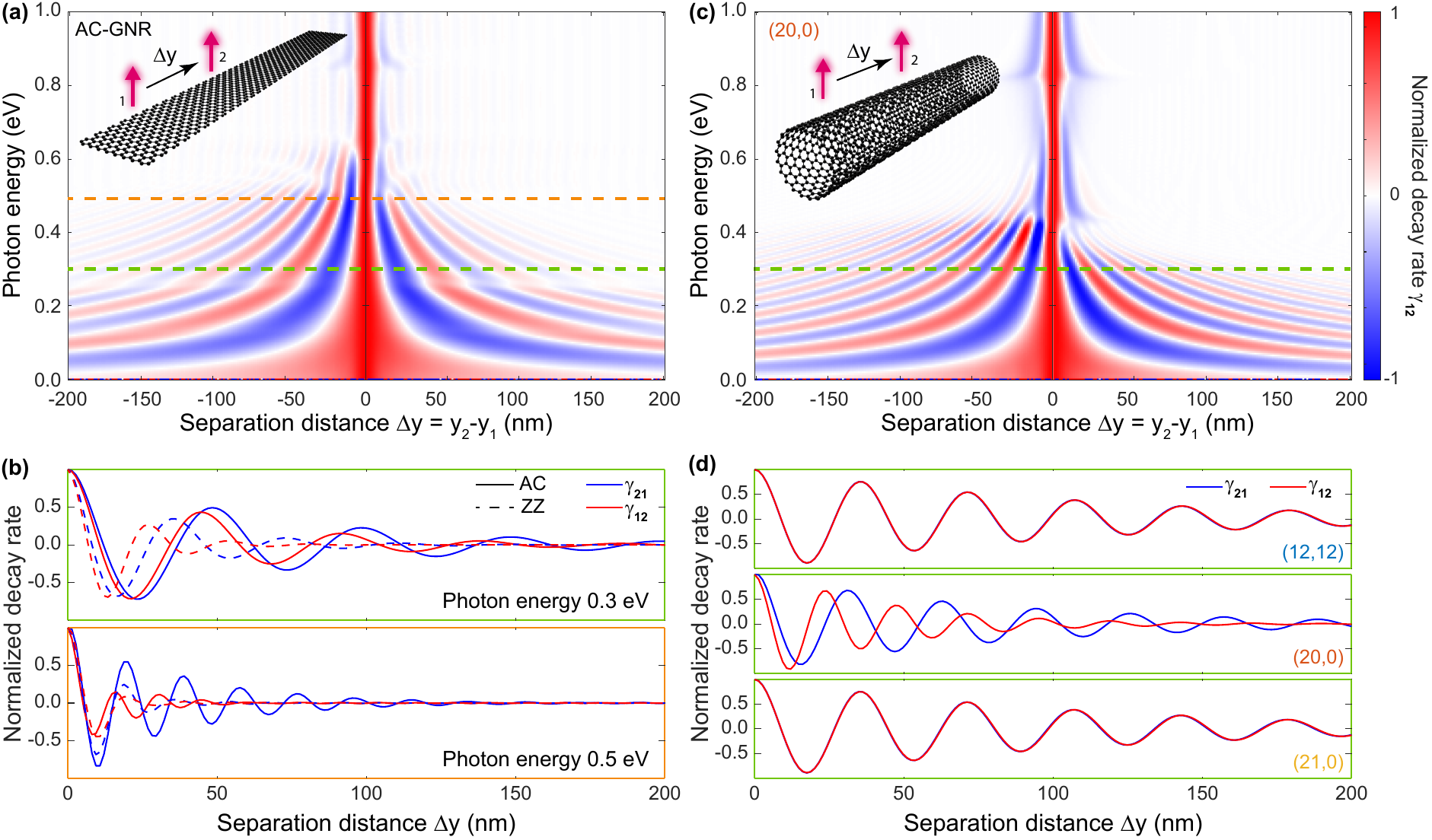}
    \caption{\textbf{Dipole-dipole interactions mediated by nonreciprocal plasmons in 1D carbon nanostructures}. (a) The normalized dipole-dipole interaction rate $\gamma_{12}$ (associated with the field produced on dipole 1 at $\rb_1$ by dipole 2 at $\rb_2$) is plotted as a a function of the dipole separation $\Delta y=y_2-y_1$ along the GNR symmetry direction $\yy$ when both dipoles are centered 5\,nm above and oriented normally to the drift-biased ($E^{\rm DC}=5.2\times10^6$\,V/m) $W\approx10$\,nm wide AC edge-terminated GNR (see inset). (b) The interaction rates $\gamma_{21}$ (blue curves) and $\gamma_{12}$ (red curves) for AC (solid curves) and ZZ (dashed curves) edge-terminated GNRs are shown for dipoles emitting at $\hbar\omega=0.3$\,eV (upper panel) and $\hbar\omega=0.5$\,eV (lower panel), i.e., at the photon energies indicated in panel (a) by the dashed horizontal lines color-coordinated with the panel outlines in (b). Panel (c) shows results analogous to those in panel (a) when the dipoles are oriented normally to and positioned 5\,nm away from the axis of a (20,0) CNT, while panel (d) compares the directional dipole-dipole interaction rates of (12,12), (20,0), and (21,0) CNTs in the upper, middle, and lower panels, respectively, in analogy to panel (b) for GNRs, when the dipoles are emitting at $\hbar\omega=0.3$\,eV, indicated by the horizontal dashed line in panel (c). All results are shown for charge carrier doping levels corresponding to $\EF=0.5$\,eV in the absence of drift currents.
    }
\label{fig:fig5}
\end{figure*}

\subsection{Dipole-dipole interaction}

Nonreciprocal photonic environments offer appealing possibilities to manipulate nonlocal interactions between light scatterers [e.g., nanoparticles or quantum light emitters (QEs)]. To explore this possibility in drift-biased carbon nanostructures, we consider the decay rate enhancement of a QE with dipole moment $\pb_1$ placed at $\rb_1$ due to the field produced by a second emitter with dipole moment $\pb_2$ located at $\rb_2$. The dipole-dipole interaction is quantified by the coupling rate $\Gamma_{12}=(2/\hbar)\int_{-\infty}^\infty{\rm d}q\Dm_q^{(12)}(x_2,z_2,\ww)$, where the analogous cross density of optical states (CDOS) function
\begin{equation} \label{eq:LDOS_12}
    \Dm_q^{(ij)} = \Imm\clpar{\pb_i^*\cdot\vec{\mathcal{E}}_{i,q}(x_j,z_j,\ww)}
\end{equation}
is expressed in terms of the Fourier components $\vec{\mathcal{E}}_{i,q}(x_j,z_j,\ww)$ of the electric field produced by $\pb_i$ at $\rb_j$. Results for interacting emitters are presented in Fig.~\ref{fig:fig5}a, where we show the normalized decay rate \cite{huidobro2012superradiance}
\begin{equation}
    \gamma_{12} = \frac{\Gamma_{12}(\Delta y)}{\Gamma_{11}}
\end{equation}
for two dipoles centered 5\,nm above an AC edge-terminated GNR and separated by a distance $\Delta y = y_2-y_1$ along the symmetry axis. The decay rate exhibits distinct behavior when the first dipole is placed in front ($\Delta y>0$) or behind ($\Delta y<0$) the second dipole (the drift current direction is set to $y>0$). In analogy to the Doppler shift, the spatial oscillation of the dipole-dipole coupling, quantified by $\gamma_{12}$, exhibits a longer or shorter period depending on the relative positions of the dipoles and drift current flow direction. We explore this phenomenon in Fig.~\ref{fig:fig5}b, where the normalized decay rates computed for AC and ZZ GNRs are shown at the indicated photon energies, also highlighted by the color-coded horizontal dashed lines in Fig.~\ref{fig:fig5}a. The atomistic calculations reveal that the AC GNR enables stronger light-matter interaction than the ZZ edge-terminated GNR, resulting in a longer range of interaction as well as a larger dissymmetry in $\gamma_{12}$.

In Fig.~\ref{fig:fig5}c,d we explore nonreciprocal dipole-dipole coupling for CNTs. In analogy to the results for the AC edge-terminated GNR shown in Fig.~\ref{fig:fig5}a, we present the spatially resolved dipole-dipole interaction rate $\gamma_{12}$ computed for the (20,0) CNT when the two interacting dipoles are each placed 5\,nm away from the CNT axis and separated by a distance $\Delta y = y_2-y_1$ (see inset). The results for the (20,0) CNT are qualitatively similar to those obtained for the AC GNR. In Fig.~\ref{fig:fig5}d, we compare the nonreciprocal dipole-dipole interaction for dipoles oscillating at $\hbar\omega=0.3$\,eV, for (12,12), (20,0), and (21,0) CNTs in the upper, middle (c.f. the horizontal dashed line in Fig.~\ref{fig:fig5}c), and lower panels. The (12,12) and (21,0) CNTs exhibit an almost fully reciprocal response, as anticipated from the minor effect of the applied drift current observed in the LDOS results of Fig.~\ref{fig:CNT}. In contrast, the moderately semiconducting (20,0) drift-biased CNT exhibits a strong nonreciprocal plasmonic response, as indicated by comparing the dipole-dipole coupling rates $\gamma_{12}$ and $\gamma_{21}$. Overall, the relative variations in the dipole-dipole interaction produced by nonreciprocal effects are substantial and can reach similar values as the dipole-dipole interaction in the reciprocal structures.

%%%%%%%%%%%%%%%%%%%%%%%%%%%%%%%%%%%%%%%%%%%%%%%%%%%%%%%%%%%%%%%%%%%%%%%%%%%%%%%%
\section{Conclusions}

We have shown that drift currents in one-dimensional carbon nanostructures can induce a large nonreciprocal response that dramatically affects the propagation of guided plasmon polaritons on nanometer length scales. Our atomistic simulations indicate that the nonreciprocal plasmonic response is highly sensitive to features in the electronic band structure of GNRs and CNTs, where a particularly large asymmetry in the response can be induced due to the presence of electronic band gaps. We further show that the nonlocal interaction between distant emitters can be manipulated by a drift current. While the applied DC field strength used to establish the drift current is commensurate with that used in experiments, we envision that qualitatively similar effects can be produced by using microwave fields, such that a drift current with a well defined direction is maintained during a fraction of the optical period in the nanosecond regime, enabling nonreciprocal optical responses over shorter times at infrared frequencies. 

%%%%%%%%%%%%%%%%%%%%%%%%%%%%%%%%%%%%%%%%%%%%%%%%%%%%%%%%%%%%%%%%%%%%%%%%%%%%%%%%
\section{Methods}

We develop models to describe the nonlocal optical response of drift-biased one-dimensional graphene nanostructures. We begin by deriving analytical expressions for the conductivity associated with free electrons in highly doped graphene, using the Boltzmann transport equation (BTE). We then outline a prescription to implement the nonlocal conductivity of drift-biased graphene in classical electrodynamic models for GNRs and CNTs. Classical models are contrasted with atomistic simulations of 1D graphene structures based on tight-binding electronic states and the RPA.

\subsection{Conductivity of drift-biased graphene}

We consider conduction electrons in a graphene sheet spanning the $\Rb=(x,y)$ plane, characterized by their parallel wave vector $\kb$ and energy $E_\kb=\pm\hbar\vF\abs{\kb}$ for electron (upper sign) and hole (lower sign) doping. In the presence of a DC field $\Eb^{\rm DC}$, the drifting of electrons at velocity $\vb$ can be described by a skewed Fermi-Dirac distribution \cite{AM1976}
\begin{equation}
    \fk^{(0)} = \left[\ee^{\ccpar{E_\kb-\mu-\hbar\vb\cdot\kb}/\kB T}+1\right]^{-1} ,
\end{equation}
where $\mu$ is the chemical potential and $T$ is the electronic temperature. The linear response to an applied optical field $\Eb$ is then given in the BTE formalism as
\begin{equation} \label{eq:BTE}
    \pd{\fk^{(1)}}{t} \pm\vF\kkb\cdot\nabla_\Rb\fk^{(1)} - \gamma\fk^{(1)} = -\frac{e}{\hbar}\Eb\cdot\nabla_\kb\fk^{(0)} ,
\end{equation}
from which the 2D current density is obtained as
\begin{equation} \label{eq:J_R}
    \Jb(\Rb,t) = -\frac{e\vF g_{\rm s}g_{\rm v}}{4\pi^2}\int{\rm d}^2\kb \, \kkb \fk^{(1)}(\Rb,t) , 
\end{equation}
where $g_{\rm s}=2$ and $g_{\rm v}=2$ are spin and valley degeneracies, respectively. Decomposing the field and associated response functions in spatial and temporal Fourier components as $\Eb(\Rb,t)=\Eb \, \ee^{\ii(\Qb\cdot\Rb-\ii\ww t)}$, the linear distribution becomes
\begin{equation}
    \fk^{(1)} = -\frac{\ii e}{\hbar}\frac{\nabla_\kb\fk^{(0)}\cdot\Eb}{\ww+\ii\gamma\mp\vF\kkb\cdot\Qb} ,
\end{equation}
which, after evaluating the gradient and inserting the result into Eq.~\eqref{eq:J_R}, leads to the current
\begin{equation} \label{eq:J_Q}
    \Jb = \frac{\ii e^2\mu_{\rm D}}{\pi^2\hbar^2}\int_0^{2\pi} {\rm d}\varphi_\kb\frac{\vF\kkb}{\ww+\ii\gamma\mp\vF\kkb\cdot\Qb}\frac{(\vF\kkb-\vb) \cdot\Eb}{(\vF-\vb\cdot\kb)^2} .
\end{equation}
Here, $\mu_{\rm D}=\mu+2\kB T\log\ccpar{1+\ee^{-\mu/\kB T}}$ is a temperature-dependent Drude weight, and the angular integral is performed over all electron wave vector directions. The current in Eq.~\eqref{eq:J_Q} implicitly defines the conductivity tensor $\sigma$ through Ohm's law $\Jb=\sigma\cdot\Eb$.

In the specific case where the drift velocity $\vb=\beta\vF\hat{\vb}$ is parallel to the in-plane optical wave vector $\Qb$, the anisotropic conductivity tensor elements for the electric field parallel ($\sigma_\parallel$) and perpendicular ($\sigma_\perp$) to the drift direction can be obtained analytically from Eq.~\eqref{eq:J_Q} and Ohm's law as
\begin{subequations}
\begin{align}
    \sigma_\parallel (Q) &= \frac{2\sigma_\ww}{(\alpha+\beta)^2}\ccpar{\frac{1+\alpha\beta}{\sqrt{1-\alpha^2}}-\sqrt{1-\beta^2}} ,  \\
    \sigma_\perp (Q) &= \frac{2\sigma_\ww}{(\alpha+\beta)^2}\ccpar{\frac{1+\alpha\beta}{\sqrt{1-\beta^2}}-\sqrt{1-\alpha^2}} ,
\end{align}
\end{subequations}
where $\sigma_\ww=\ii e^2 \mu_{\rm D}/\pi\hbar^2(\ww+\ii\gamma)$ is the local Drude conductivity of graphene and $\alpha=\mp\vF Q/(\ww+\ii\gamma)$. In what follows, we employ the above nonlocal conductivity of drift-biased graphene to perform classical electrodynamic calculations of GNRs and CNTs.

The temperature- and drift-dependent chemical potential $\mu(\beta)$ is determined by imposing conservation of electron population such that 
\begin{equation} \label{eq:Ne}
    N_e = 4 \sum_\kb f^{(0)}_\kb(\mu,\beta) .
\end{equation}
The chemical potential $\mu(\beta)$ depends on the bias electric DC field $\Eb^{\rm DC}$, and $N_e$ is fixed by imposing the condition $\mu=\EF$ when $\beta=0$. The factor of 4 includes the spin and valley degeneracies (see Fig.~S4a in the SI). 

\subsection{Classical description of graphene nanoribbons (GNRs)}

In the quasistatic approximation, the optical response of a 2D system to an external potential $\Phi^{\rm ext}$ is characterized by the self-consistent scalar potential
\begin{equation} \label{eq:Phi_r}
    \Phi(\rb) = \Phiext(\rb) + \frac{1}{\epseff}\int{\rm d}^2\Rb'\frac{\rhoind(\Rb')}{\abs{\rb-\Rb'}} ,
\end{equation}
where the Coulomb interaction has been corrected with the average permittivity $\epseff\equiv(\epsa + \epsb)/2$ to account for the effect of screening by the homogeneous dielectric media above ($\epsa$) and below ($\epsb$) the 2D material, and $\rhoind(\Rb)$ is the induced 2D charge density in the material, placed in the $z=0$ plane. Invoking the continuity equation $\rhoind = -(\ii/\ww)\nabla\cdot\Jb$, Ohm's law $\Jb=\sigma\cdot\Eb$, and the electrostatic relation $\Eb=-\nabla\Phi$, the potential at points within the 2D material can be expressed self-consistently as
\begin{equation} \label{eq:Phi_Phiext}
    %\Phi(\Rb) = \Phiext(\Rb) + \frac{\ii}{\epseff\ww}\Vm\Dm\Phi(\Rb) ,
    \ccpar{1-\frac{\ii}{\epseff\ww}\Mm}\Phi(\Rb) = \Phiext(\Rb) ,
    %\Phi(\Rb) = \Phiext(\Rb) + \frac{\ii}{\epseff\ww}\Mm\Phi(\Rb) ,
\end{equation}
where
\begin{equation}
    \Mm\Phi(\Rb) = \int{\rm d}^2\Rb'\frac{1}{\abs{\Rb-\Rb'}}\nabla_{\Rb'}\cdot\sqpar{\sigma(\Rb')\cdot\nabla_{\Rb'}\Phi(\Rb')}
\end{equation}
is a linear integro-differential operator. Here, we are interested in studying the local photonic density of states in the vicinity of the 2D structure, characterized by the induced field generated by a point dipole $\pb$ at $\rb_0$ acting back on itself, that is,
\begin{equation} \label{Eq:e_ind_ribb}
    \Eb^{\rm ind}(\rb_0) = -\frac{1}{\epseff} \nabla_{\rb_0}\int{\rm d}^2\Rb'\frac{\rhoind(\Rb')}{\abs{\rb_0-\Rb'}} ,
    %\Eb^{\rm ind}(\rb_0) = -\frac{1}{\epseff}\Mm\Phi(\rb_0) ,
\end{equation}
where $\rhoind(\Rb)=(\ii/\ww)\nabla_\Rb\cdot\sqpar{\sigma(\Rb)\cdot\nabla_\Rb\Phi(\Rb)}$ is the induced charge produced in the ribbon by the external potential $\Phiext(\rb)= \eps_a^{-1} \, \pb\cdot\nabla_{\rb_0}\left\vert\rb-\rb_0\right\vert^{-1}$.

For a nanoribbon of finite width $W$ along $\xx$ and infinite extension along $\yy$, the scalar potentials are conveniently decomposed in plane waves with wave vector $q$ as $\Phi(\Rb)= (2\pi)^{-1} \int {\rm d}q \, \phi_q(x) \, \ee^{\ii q y}$, such that the spatial Fourier components satisfy the relation
\begin{equation} \label{eq:phi_q}
    %\phi_q(x) = \phiext_q(x) + \frac{\ii}{\epseff\ww} \Mm_q\phi_q(x) ,
    \ccpar{1-\frac{\ii}{\epseff\ww}\Mm_q}\phi_q(x) = \phiext_q(x) ,
\end{equation}
where we use the integro-differential operator
\begin{align}
    \Mm_q\phi_q(x) &= 2\int{\rm d}x' K_0\ccpar{\abs{q}\abs{x-x'}} \\
    &\times \left\{
        \partial_{x'}\sqpar{\sigma_\perp(x')\partial_{x'}\phi_q(x')} 
        -q^2 \sigma_\parallel(x') \phi_q(x')
    \right\} \nonumber
\end{align}
and
\begin{equation}
    \phiext_q(x) = \frac{2}{\epsa} \pb \cdot \nabla_{\rb_0}\clpar{K_0\sqpar{ |q| \sqrt{(x-x_0)^2+z_0^2}}\ee^{-\ii q y_0}}
\end{equation}
is the Fourier transform of the external potential. Following the prescription of Refs.~\cite{christensen2015kerr,rasmussen2023nonlocal}, Eq.~\eqref{eq:phi_q} is solved by discretizing the spatial coordinate $x$ in the ribbon domain defined through the conductivity as $\sigma(x)=f(x)\sigma(\ww)$, where $f(x)$ is a scalar function that is unity for points inside the ribbon and vanishes elsewhere, so that the potential can be obtained through matrix inversion. From Eq.~\eqref{Eq:e_ind_ribb}, the induced field acting back on the dipole at $\rb_0$ can then be found by evaluating
\begin{equation} \label{eq:E_ind}
    \Eb^{\rm ind}(\rb_0) = \int_{-\infty}^\infty\frac{{\rm d}q}{2\pi}\vec{\mathcal{E}}_q(\rb_0) ,
\end{equation}
where
\begin{align}
    \vec{\mathcal{E}}_q(\rb_0) &= -\frac{\ii \ee^{\ii q y_0}}{\epseff\ww}\nabla_{\rb_0} \int {\rm d}x'K_0\sqpar{|q|\sqrt{(x_0-x')^2+z_0^2}} \nonumber \\
    &\times \left\{
        \partial_{x'}\sqpar{\sigma_\perp(x')\partial_{x'}\phi_q(x')} 
        -q^2 \sigma_\parallel(x') \phi_q(x')
    \right\}
\end{align}
is the induced electric field decomposed in wave vectors $q$, which we use in Eq.~\eqref{eq:LDOS_q} to compute the LDOS.

\subsection{Classical description of carbon nanotubes (CNTs)}

Following the procedure described in Ref.~\cite{martinmoreno2015ultraefficient}, the induced electric field produced by a point dipole $\pb$ in the presence of a CNT of radius $a$ can be written as
\begin{equation} \label{Eq:E_field_cylinder}
    \Eb^{\rm ind}(\rb) =  -\nabla_\rb \left.\left[ (\pb\cdot\nabla_\rb)W^{\rm ind} (\rb,\rb_0,\omega) \right] \right\vert_{\rb_0=\rb} ,
\end{equation}
where $W^{\rm ind}$ denotes the induced part of the total screened interaction $W(\rb,\rb_0) = |\rb-\rb_0|^{-1}+W^{\rm ind}(\rb,\rb_0)$. The function $W$ is defined as the electric scalar potential created at $\rb$ by an oscillating point charge $\ee^{-\ii \omega t}$ located at $\rb_0$, and is obtained by solving Poisson’s equation. For $\rb$ and $\rb_0$ both outside the cylinder, $W^{\rm ind}$ adopts the following analytical form:
\begin{align} 
    &W^{\rm ind} \left(\rb, \rb_0, \omega\right)=  \frac{2}{\pi} \sum_{m=0}^{\infty}\left(2-\delta_{m 0}\right) \cos \left[m\left(\varphi-\varphi_0\right)\right] \nonumber \\ 
    & \times \int_{-\infty}^{\infty} {\rm d}q \, r_m(q) \, \ee^{\ii q (z-z_0)} K_m(|q|\rho) K_m\left(|q| \rho_0\right) ,
\end{align}
where
\begin{equation} \label{Eq:reflection_coefficient}
    r_m(q) = -\frac{I_m^2(|q|a) \Delta}{K_m(|q|a) I_m(|q|a) \Delta + 1}
\end{equation}
is a reflection coefficient expressed in terms of the quantity
\begin{equation}
    \Delta(q) = \frac{4\pi\ii}{\omega a}\left[m^2\sigma_{\perp}(q) + (qa)^2\sigma_{\parallel}(q)\right] .
\end{equation}
Using the screened interaction and taking the gradient of the potential, the wave-vector components of the induced electric field [see Eq.~\eqref{eq:E_ind}] are found to be
\begin{equation} \label{eq:Eq_CNT}
    \vec{\mathcal{E}}_q(\rb_0) = 2 \hat{\rho} \sum_{m=0}^{\infty} \left(\delta_{m 0}-2\right)  \, q^2\, r_m(q) \left[K_m'(|q| \rho_0) \right]^2 
\end{equation}
for a unit dipole oriented along the $\hat{\rho}$ direction and placed at the position of the dipole $\rb_0=(\rho_0,0,0)$. The above expression is used in Eq.~\eqref{eq:LDOS_q} to compute the direction-dependent LDOS.

\subsection{Atomistic description of one-dimensional carbon nanostructures}

Following the approach of Refs.~\cite{thongrattanasiri2012quantum,cox2016quantum}, the electrons in carbon p$_z$ orbitals $\ket{l,m}$ (i.e., located at the carbon atom positions $\Rb_l$ in each unit cell $m$ of a one-dimensional GNR or CNT) are described as Bloch states
\begin{equation}\label{eq:jk_2_lm}
    \ket{j,k} = \frac{1}{\sqrt{M}} \sum_{l,m} a_{jl,k}\ee^{\ii kmb}\ket{l,m}.
\end{equation}
Here, $M$ denotes a (large) number of unit cells, $k$ is the Bloch wave vector in the direction of translational lattice invariance, $b$ is the unit cell period, and the coefficients $a_{jl,k}$ are determined by diagonalizing the tight-binding Hamiltonian $\Hm_{ll',k}^{\rm TB}$ as
\begin{equation}
    \sum_{l'}\Hm_{ll',k}^{\rm TB}a_{jl',k} = \hbar\vep_{j,k}a_{jl,k}
\end{equation}
to yield the energies $\hbar\vep_{j,k}$ of electrons in each band $j$ as a function of Bloch wave vector $k$ in the direction of translational invariance. 

In analogy to Eqs.~\eqref{eq:Phi_r} and \eqref{eq:phi_q}, we write the spatial Fourier components of the self-consistent potential at site $l$ produced in response to an external potential $\phiext_{q,l}\ee^{\ii q y}$ as
\begin{equation}
    \phi_{q,l} = \phiext_{q,l} + \sum_{l'}v_{q,ll'}\rhoind_{q,l} ,
\end{equation}
where $v_{q,ll'} = \sum_m v_{l0,l'm} \ee^{-\ii q m b}$ is the Fourier transform of $v_{l0,l'm}$ (i.e., the spatial dependence of the Coulomb interaction between electrons in orbitals $l$ and $l'$ separated by $m$ unit cells), and 
\begin{equation}
    \rhoind_{q,l} = \sum_{l'} \chi_{q,ll'}^0 \phi_{q,l'}
\end{equation}
is the induced charge, which, in the framework of the RPA, is expressed in terms of the non interacting susceptibility 
\begin{align}
    \chi_{q,ll'}^0 = &\frac{2e^2}{\hbar}\frac{b}{2\pi}\int_{-\pi/b}^{\pi/b}{\rm d}k \sum_{jj'}\ccpar{f_{j',k-q}-f_{j,k}} \nonumber \\
    &\times\frac{a_{jl,k}a_{j'l,k-q}^*a_{jl',k}^*a_{j'l',k-q}}{\ww+\ii/(2\tau)-\ccpar{\vep_{j,k}-\vep_{j',k-q}}} , \label{eq:chi0}
\end{align}
comprising a sum over electronic transitions weighted by occupation factors $f_{j,k}$. 

In the absence of an applied DC field, the occupation factors entering Eq.~\eqref{eq:chi0} are determined by the Fermi-Dirac statistics, that is,
\begin{equation}
    f_{j,k} = f_{j,k}^{(0)} \equiv \left[\ee^{(\hbar\vep_{j,k}-\mu)/\kB T}+1\right]^{-1} . 
\end{equation}
To incorporate the applied DC field in atomistic calculations that capture the full one-dimensional multi-band electronic structure of GNRs and CNTs, we obtain $f_{j,k}$ by solving the Boltzmann transport equation [see Eq.~\eqref{eq:BTE}] in the static limit:
\begin{equation} \label{Eq:f_grarpa}
    \ccpar{1-\frac{e\tau}{\hbar}E^{\rm DC}\pd{}{k}}f_{j,k} = f_{j,k}^{(0)},
\end{equation}
which is the same as Eq.~\eqref{eq:f_E_DC} in the main text. Note that the chemical potential is chosen to preserve the number of electrons [see Eq.~\eqref{eq:Ne}].

For a specified in-plane field $\Eb^{\rm DC}$, the electron drift velocity is estimated from $v = \mu^{\rm DC} E^{\rm DC}$, where $\mu^{\rm DC}$ is the  electron mobility, which dictates the mean-free path via \cite{garciadeabajo2014graphene}
\begin{equation}
    \tau = \frac{\mu^{\rm DC} \EF}{e \vF^2}
\end{equation}
and is related to the damping as $\gamma=\tau^{-1}$ (see Fig.~S4b in the SI for different choices of parameters). The above expressions thus relate the parameter $\beta=v/\vF$ entering the semiclassical calculations to the applied static field $E^{\rm DC}$ used in the TB-RPA formalism. In the SI, we study the electronic band structures of AC and ZZ GNRs in Figs.~S5 and S6, respectively, as well as the occupation factors $f_{j,k}$, for different configurations of charge carrier doping and applied fields $E^{\rm DC}$. Similarly, we show in Fig.~S7 the electronic bands for the CNTs of $\approx1.6$\,nm diameter studied in this work.

\section*{Acknowledgments}

The authors thank N.~M.~R.~Peres for fruitful discussions.
A.R.E. and F.J.G.A. acknowledge support from ERC (101141220-QUEFES) and the Spanish MICINN (Severo Ochoa CEX2019-000910-S).
J.D.C. acknowledges support from Independent Research Fund Denmark (Grant No.~0165-00051B).
The Center for Polariton-driven Light--Matter Interactions (POLIMA) is funded by the Danish National Research Foundation (Project No.~DNRF165).

%\bibliographystyle{apsrev4-2}
%\bibliography{refs}

\begin{thebibliography}{50}%
\makeatletter
\providecommand \@ifxundefined [1]{%
 \@ifx{#1\undefined}
}%
\providecommand \@ifnum [1]{%
 \ifnum #1\expandafter \@firstoftwo
 \else \expandafter \@secondoftwo
 \fi
}%
\providecommand \@ifx [1]{%
 \ifx #1\expandafter \@firstoftwo
 \else \expandafter \@secondoftwo
 \fi
}%
\providecommand \natexlab [1]{#1}%
\providecommand \enquote  [1]{``#1''}%
\providecommand \bibnamefont  [1]{#1}%
\providecommand \bibfnamefont [1]{#1}%
\providecommand \citenamefont [1]{#1}%
\providecommand \href@noop [0]{\@secondoftwo}%
\providecommand \href [0]{\begingroup \@sanitize@url \@href}%
\providecommand \@href[1]{\@@startlink{#1}\@@href}%
\providecommand \@@href[1]{\endgroup#1\@@endlink}%
\providecommand \@sanitize@url [0]{\catcode `\\12\catcode `\$12\catcode
  `\&12\catcode `\#12\catcode `\^12\catcode `\_12\catcode `\%12\relax}%
\providecommand \@@startlink[1]{}%
\providecommand \@@endlink[0]{}%
\providecommand \url  [0]{\begingroup\@sanitize@url \@url }%
\providecommand \@url [1]{\endgroup\@href {#1}{\urlprefix }}%
\providecommand \urlprefix  [0]{URL }%
\providecommand \Eprint [0]{\href }%
\providecommand \doibase [0]{https://doi.org/}%
\providecommand \selectlanguage [0]{\@gobble}%
\providecommand \bibinfo  [0]{\@secondoftwo}%
\providecommand \bibfield  [0]{\@secondoftwo}%
\providecommand \translation [1]{[#1]}%
\providecommand \BibitemOpen [0]{}%
\providecommand \bibitemStop [0]{}%
\providecommand \bibitemNoStop [0]{.\EOS\space}%
\providecommand \EOS [0]{\spacefactor3000\relax}%
\providecommand \BibitemShut  [1]{\csname bibitem#1\endcsname}%
\let\auto@bib@innerbib\@empty
%</preamble>
\bibitem [{\citenamefont {Hohenester}(2020)}]{hohenester2020nano}%
  \BibitemOpen
  \bibfield  {author} {\bibinfo {author} {\bibfnamefont {U.}~\bibnamefont
  {Hohenester}},\ }\href@noop {} {\emph {\bibinfo {title} {Nano and Quantum
  Optics}}}\ (\bibinfo  {publisher} {Springer},\ \bibinfo {year}
  {2020})\BibitemShut {NoStop}%
\bibitem [{\citenamefont {Potton}(2004)}]{potton2004reciprocity}%
  \BibitemOpen
  \bibfield  {author} {\bibinfo {author} {\bibfnamefont {R.~J.}\ \bibnamefont
  {Potton}},\ }\href {https://doi.org/10.1088/0034-4885/67/5/R03} {\bibfield
  {journal} {\bibinfo  {journal} {Rep.\ Prog.\ Phys.}\ }\textbf {\bibinfo
  {volume} {67}},\ \bibinfo {pages} {717} (\bibinfo {year} {2004})}\BibitemShut
  {NoStop}%
\bibitem [{\citenamefont {Chin}\ \emph {et~al.}(2013)\citenamefont {Chin},
  \citenamefont {Steinle}, \citenamefont {Wehlus}, \citenamefont {Dregely},
  \citenamefont {Weiss}, \citenamefont {Belotelov}, \citenamefont {Stritzker},\
  and\ \citenamefont {Giessen}}]{chin2013nonreciprocal}%
  \BibitemOpen
  \bibfield  {author} {\bibinfo {author} {\bibfnamefont {J.~Y.}\ \bibnamefont
  {Chin}}, \bibinfo {author} {\bibfnamefont {T.}~\bibnamefont {Steinle}},
  \bibinfo {author} {\bibfnamefont {T.}~\bibnamefont {Wehlus}}, \bibinfo
  {author} {\bibfnamefont {D.}~\bibnamefont {Dregely}}, \bibinfo {author}
  {\bibfnamefont {T.}~\bibnamefont {Weiss}}, \bibinfo {author} {\bibfnamefont
  {V.~I.}\ \bibnamefont {Belotelov}}, \bibinfo {author} {\bibfnamefont
  {B.}~\bibnamefont {Stritzker}},\ and\ \bibinfo {author} {\bibfnamefont
  {H.}~\bibnamefont {Giessen}},\ }\href {https://doi.org/10.1038/ncomms2609}
  {\bibfield  {journal} {\bibinfo  {journal} {Nat.\ Commun.}\ }\textbf
  {\bibinfo {volume} {4}},\ \bibinfo {pages} {1599} (\bibinfo {year}
  {2013})}\BibitemShut {NoStop}%
\bibitem [{\citenamefont {Davoyan}\ and\ \citenamefont
  {Engheta}(2014)}]{davoyan2014electrically}%
  \BibitemOpen
  \bibfield  {author} {\bibinfo {author} {\bibfnamefont {A.}~\bibnamefont
  {Davoyan}}\ and\ \bibinfo {author} {\bibfnamefont {N.}~\bibnamefont
  {Engheta}},\ }\href {https://doi.org/10.1038/ncomms6250} {\bibfield
  {journal} {\bibinfo  {journal} {Nat.\ Commun.}\ }\textbf {\bibinfo {volume}
  {5}},\ \bibinfo {pages} {5250} (\bibinfo {year} {2014})}\BibitemShut
  {NoStop}%
\bibitem [{\citenamefont {Yu}\ \emph {et~al.}(2008)\citenamefont {Yu},
  \citenamefont {Veronis}, \citenamefont {Wang},\ and\ \citenamefont
  {Fan}}]{yu2008oneway}%
  \BibitemOpen
  \bibfield  {author} {\bibinfo {author} {\bibfnamefont {Z.}~\bibnamefont
  {Yu}}, \bibinfo {author} {\bibfnamefont {G.}~\bibnamefont {Veronis}},
  \bibinfo {author} {\bibfnamefont {Z.}~\bibnamefont {Wang}},\ and\ \bibinfo
  {author} {\bibfnamefont {S.}~\bibnamefont {Fan}},\ }\href
  {https://doi.org/10.1103/PhysRevLett.100.023902} {\bibfield  {journal}
  {\bibinfo  {journal} {Phys.\ Rev.\ Lett.}\ }\textbf {\bibinfo {volume}
  {100}},\ \bibinfo {pages} {023902} (\bibinfo {year} {2008})}\BibitemShut
  {NoStop}%
\bibitem [{\citenamefont {Bliokh}\ \emph {et~al.}(2018)\citenamefont {Bliokh},
  \citenamefont {Rodr{\'\i}guez-Fortu{\~n}o}, \citenamefont {Bekshaev},
  \citenamefont {Kivshar},\ and\ \citenamefont {Nori}}]{bliokh2018electric}%
  \BibitemOpen
  \bibfield  {author} {\bibinfo {author} {\bibfnamefont {K.~Y.}\ \bibnamefont
  {Bliokh}}, \bibinfo {author} {\bibfnamefont {F.~J.}\ \bibnamefont
  {Rodr{\'\i}guez-Fortu{\~n}o}}, \bibinfo {author} {\bibfnamefont {A.~Y.}\
  \bibnamefont {Bekshaev}}, \bibinfo {author} {\bibfnamefont {Y.~S.}\
  \bibnamefont {Kivshar}},\ and\ \bibinfo {author} {\bibfnamefont
  {F.}~\bibnamefont {Nori}},\ }\href {https://doi.org/10.1364/OL.43.000963}
  {\bibfield  {journal} {\bibinfo  {journal} {Opt.\ Lett.}\ }\textbf {\bibinfo
  {volume} {43}},\ \bibinfo {pages} {963} (\bibinfo {year} {2018})}\BibitemShut
  {NoStop}%
\bibitem [{\citenamefont {Li}\ \emph {et~al.}(2024)\citenamefont {Li},
  \citenamefont {Tsakmakidis}, \citenamefont {Jiang}, \citenamefont {Shen},
  \citenamefont {Zhang}, \citenamefont {Yan}, \citenamefont {Sun},\ and\
  \citenamefont {Shen}}]{li2024unidirectional}%
  \BibitemOpen
  \bibfield  {author} {\bibinfo {author} {\bibfnamefont {S.}~\bibnamefont
  {Li}}, \bibinfo {author} {\bibfnamefont {K.~L.}\ \bibnamefont {Tsakmakidis}},
  \bibinfo {author} {\bibfnamefont {T.}~\bibnamefont {Jiang}}, \bibinfo
  {author} {\bibfnamefont {Q.}~\bibnamefont {Shen}}, \bibinfo {author}
  {\bibfnamefont {H.}~\bibnamefont {Zhang}}, \bibinfo {author} {\bibfnamefont
  {J.}~\bibnamefont {Yan}}, \bibinfo {author} {\bibfnamefont {S.}~\bibnamefont
  {Sun}},\ and\ \bibinfo {author} {\bibfnamefont {L.}~\bibnamefont {Shen}},\
  }\href {https://doi.org/10.1038/s41467-024-50287-z} {\bibfield  {journal}
  {\bibinfo  {journal} {Nat.\ Commun.}\ }\textbf {\bibinfo {volume} {15}},\
  \bibinfo {pages} {5992} (\bibinfo {year} {2024})}\BibitemShut {NoStop}%
\bibitem [{\citenamefont {{Hassani Gangaraj}}\ and\ \citenamefont
  {Monticone}(2022)}]{hassani2022drifting}%
  \BibitemOpen
  \bibfield  {author} {\bibinfo {author} {\bibfnamefont {S.~A.}\ \bibnamefont
  {{Hassani Gangaraj}}}\ and\ \bibinfo {author} {\bibfnamefont
  {F.}~\bibnamefont {Monticone}},\ }\href
  {https://doi.org/10.1021/acsphotonics.1c01294} {\bibfield  {journal}
  {\bibinfo  {journal} {ACS Photonics}\ }\textbf {\bibinfo {volume} {9}},\
  \bibinfo {pages} {806} (\bibinfo {year} {2022})}\BibitemShut {NoStop}%
\bibitem [{\citenamefont {Liu}\ \emph {et~al.}(2022)\citenamefont {Liu},
  \citenamefont {Guo}, \citenamefont {Li},\ and\ \citenamefont
  {Fan}}]{liu2022thermal}%
  \BibitemOpen
  \bibfield  {author} {\bibinfo {author} {\bibfnamefont {T.}~\bibnamefont
  {Liu}}, \bibinfo {author} {\bibfnamefont {C.}~\bibnamefont {Guo}}, \bibinfo
  {author} {\bibfnamefont {W.}~\bibnamefont {Li}},\ and\ \bibinfo {author}
  {\bibfnamefont {S.}~\bibnamefont {Fan}},\ }\href
  {https://doi.org/10.1186/s43593-022-00025-z} {\bibfield  {journal} {\bibinfo
  {journal} {ELight}\ }\textbf {\bibinfo {volume} {2}},\ \bibinfo {pages} {25}
  (\bibinfo {year} {2022})}\BibitemShut {NoStop}%
\bibitem [{\citenamefont {Jalas}\ \emph {et~al.}(2013)\citenamefont {Jalas},
  \citenamefont {Petrov}, \citenamefont {Eich}, \citenamefont {Freude},
  \citenamefont {Fan}, \citenamefont {Yu}, \citenamefont {Baets}, \citenamefont
  {Popovi\'{c}}, \citenamefont {Melloni}, \citenamefont {Joannopoulos} \emph
  {et~al.}}]{jalas2013what}%
  \BibitemOpen
  \bibfield  {author} {\bibinfo {author} {\bibfnamefont {D.}~\bibnamefont
  {Jalas}}, \bibinfo {author} {\bibfnamefont {A.}~\bibnamefont {Petrov}},
  \bibinfo {author} {\bibfnamefont {M.}~\bibnamefont {Eich}}, \bibinfo {author}
  {\bibfnamefont {W.}~\bibnamefont {Freude}}, \bibinfo {author} {\bibfnamefont
  {S.}~\bibnamefont {Fan}}, \bibinfo {author} {\bibfnamefont {Z.}~\bibnamefont
  {Yu}}, \bibinfo {author} {\bibfnamefont {R.}~\bibnamefont {Baets}}, \bibinfo
  {author} {\bibfnamefont {M.}~\bibnamefont {Popovi\'{c}}}, \bibinfo {author}
  {\bibfnamefont {A.}~\bibnamefont {Melloni}}, \bibinfo {author} {\bibfnamefont
  {J.~D.}\ \bibnamefont {Joannopoulos}}, \emph {et~al.},\ }\href
  {https://doi.org/10.1038/nphoton.2013.185} {\bibfield  {journal} {\bibinfo
  {journal} {Nat.\ Photonics}\ }\textbf {\bibinfo {volume} {7}},\ \bibinfo
  {pages} {579} (\bibinfo {year} {2013})}\BibitemShut {NoStop}%
\bibitem [{\citenamefont {Boroviks}\ \emph {et~al.}(2023)\citenamefont
  {Boroviks}, \citenamefont {Kiselev}, \citenamefont {Achouri},\ and\
  \citenamefont {Martin}}]{boroviks2023demonstration}%
  \BibitemOpen
  \bibfield  {author} {\bibinfo {author} {\bibfnamefont {S.}~\bibnamefont
  {Boroviks}}, \bibinfo {author} {\bibfnamefont {A.}~\bibnamefont {Kiselev}},
  \bibinfo {author} {\bibfnamefont {K.}~\bibnamefont {Achouri}},\ and\ \bibinfo
  {author} {\bibfnamefont {O.~J.~F.}\ \bibnamefont {Martin}},\ }\href
  {https://doi.org/10.1021/acs.nanolett.3c00367} {\bibfield  {journal}
  {\bibinfo  {journal} {Nano Lett.}\ }\textbf {\bibinfo {volume} {23}},\
  \bibinfo {pages} {3362} (\bibinfo {year} {2023})}\BibitemShut {NoStop}%
\bibitem [{\citenamefont {Shi}\ \emph {et~al.}(2015)\citenamefont {Shi},
  \citenamefont {Yu},\ and\ \citenamefont {Fan}}]{shi2015limitations}%
  \BibitemOpen
  \bibfield  {author} {\bibinfo {author} {\bibfnamefont {Y.}~\bibnamefont
  {Shi}}, \bibinfo {author} {\bibfnamefont {Z.}~\bibnamefont {Yu}},\ and\
  \bibinfo {author} {\bibfnamefont {S.}~\bibnamefont {Fan}},\ }\href
  {https://doi.org/10.1038/nphoton.2015.79} {\bibfield  {journal} {\bibinfo
  {journal} {Nat.\ Photonics}\ }\textbf {\bibinfo {volume} {9}},\ \bibinfo
  {pages} {388} (\bibinfo {year} {2015})}\BibitemShut {NoStop}%
\bibitem [{\citenamefont {Mahmoud}\ \emph {et~al.}(2015)\citenamefont
  {Mahmoud}, \citenamefont {Davoyan},\ and\ \citenamefont
  {Engheta}}]{mahmoud2015all}%
  \BibitemOpen
  \bibfield  {author} {\bibinfo {author} {\bibfnamefont {A.~M.}\ \bibnamefont
  {Mahmoud}}, \bibinfo {author} {\bibfnamefont {A.~R.}\ \bibnamefont
  {Davoyan}},\ and\ \bibinfo {author} {\bibfnamefont {N.}~\bibnamefont
  {Engheta}},\ }\href {https://doi.org/10.1038/ncomms9359} {\bibfield
  {journal} {\bibinfo  {journal} {Nat.\ Commun.}\ }\textbf {\bibinfo {volume}
  {6}},\ \bibinfo {pages} {8359} (\bibinfo {year} {2015})}\BibitemShut
  {NoStop}%
\bibitem [{\citenamefont {Papaj}\ and\ \citenamefont
  {Lewandowski}(2020)}]{papaj2020plasmonic}%
  \BibitemOpen
  \bibfield  {author} {\bibinfo {author} {\bibfnamefont {M.}~\bibnamefont
  {Papaj}}\ and\ \bibinfo {author} {\bibfnamefont {C.}~\bibnamefont
  {Lewandowski}},\ }\href {https://doi.org/10.1103/PhysRevLett.125.066801}
  {\bibfield  {journal} {\bibinfo  {journal} {Phys.\ Rev.\ Lett.}\ }\textbf
  {\bibinfo {volume} {125}},\ \bibinfo {pages} {066801} (\bibinfo {year}
  {2020})}\BibitemShut {NoStop}%
\bibitem [{\citenamefont {Monticone}\ \emph {et~al.}(2025)\citenamefont
  {Monticone}, \citenamefont {Mortensen}, \citenamefont
  {Fern\'{a}ndez-Dom\'{i}nguez}, \citenamefont {Luo}, \citenamefont {Zheng},
  \citenamefont {Tserkezis}, \citenamefont {Khurgin}, \citenamefont
  {Shahbazyan}, \citenamefont {Chaves}, \citenamefont {Peres}, \citenamefont
  {Wegner}, \citenamefont {Busch}, \citenamefont {Hu}, \citenamefont {Sala},
  \citenamefont {Zhang}, \citenamefont {Cirac\`{i}}, \citenamefont {Aizpurua},
  \citenamefont {Babaze}, \citenamefont {Borisov}, \citenamefont {Chen},
  \citenamefont {Christensen}, \citenamefont {Yan}, \citenamefont {Yang},
  \citenamefont {Hohenester}, \citenamefont {Huber}, \citenamefont {Wubs},
  \citenamefont {Liberato}, \citenamefont {Gon\c{c}alves}, \citenamefont
  {{Garc\'{\i}a de Abajo}}, \citenamefont {Hess}, \citenamefont {Tarasenko},
  \citenamefont {Cox}, \citenamefont {Jelver}, \citenamefont {Dias},
  \citenamefont {S\'{a}nchez}, \citenamefont {Margetis}, \citenamefont
  {G\'{o}mez-Santos}, \citenamefont {Vasilevskiy}, \citenamefont {Stauber},
  \citenamefont {Tretyakov}, \citenamefont {Simovski}, \citenamefont
  {Pakniyat}, \citenamefont {G\'{o}mez-D\'{i}az}, \citenamefont {Bondarev},
  \citenamefont {Biehs}, \citenamefont {Boltasseva}, \citenamefont {Shalaev},
  \citenamefont {Krasavin}, \citenamefont {Zayats}, \citenamefont {Al\`{u}},
  \citenamefont {Song}, \citenamefont {Brongersma}, \citenamefont {Levy},
  \citenamefont {Long}, \citenamefont {Guo}, \citenamefont {Fan}, \citenamefont
  {Bozhevolnyi}, \citenamefont {Overvig}, \citenamefont {Prud\^{e}ncio},
  \citenamefont {Silveirinha}, \citenamefont {Gangaraj}, \citenamefont
  {Argyropoulos}, \citenamefont {Huidobro}, \citenamefont {Galiffi},
  \citenamefont {Yang}, \citenamefont {Pendry},\ and\ \citenamefont
  {Miller}}]{monticone2025nonlocality}%
  \BibitemOpen
  \bibfield  {author} {\bibinfo {author} {\bibfnamefont {F.}~\bibnamefont
  {Monticone}}, \bibinfo {author} {\bibfnamefont {N.~A.}\ \bibnamefont
  {Mortensen}}, \bibinfo {author} {\bibfnamefont {A.~I.}\ \bibnamefont
  {Fern\'{a}ndez-Dom\'{i}nguez}}, \bibinfo {author} {\bibfnamefont
  {Y.}~\bibnamefont {Luo}}, \bibinfo {author} {\bibfnamefont {X.}~\bibnamefont
  {Zheng}}, \bibinfo {author} {\bibfnamefont {C.}~\bibnamefont {Tserkezis}},
  \bibinfo {author} {\bibfnamefont {J.~B.}\ \bibnamefont {Khurgin}}, \bibinfo
  {author} {\bibfnamefont {T.~V.}\ \bibnamefont {Shahbazyan}}, \bibinfo
  {author} {\bibfnamefont {A.~J.}\ \bibnamefont {Chaves}}, \bibinfo {author}
  {\bibfnamefont {N.~M.~R.}\ \bibnamefont {Peres}}, \bibinfo {author}
  {\bibfnamefont {G.}~\bibnamefont {Wegner}}, \bibinfo {author} {\bibfnamefont
  {K.}~\bibnamefont {Busch}}, \bibinfo {author} {\bibfnamefont
  {H.}~\bibnamefont {Hu}}, \bibinfo {author} {\bibfnamefont {F.~D.}\
  \bibnamefont {Sala}}, \bibinfo {author} {\bibfnamefont {P.}~\bibnamefont
  {Zhang}}, \bibinfo {author} {\bibfnamefont {C.}~\bibnamefont {Cirac\`{i}}},
  \bibinfo {author} {\bibfnamefont {J.}~\bibnamefont {Aizpurua}}, \bibinfo
  {author} {\bibfnamefont {A.}~\bibnamefont {Babaze}}, \bibinfo {author}
  {\bibfnamefont {A.~G.}\ \bibnamefont {Borisov}}, \bibinfo {author}
  {\bibfnamefont {X.-W.}\ \bibnamefont {Chen}}, \bibinfo {author}
  {\bibfnamefont {T.}~\bibnamefont {Christensen}}, \bibinfo {author}
  {\bibfnamefont {W.}~\bibnamefont {Yan}}, \bibinfo {author} {\bibfnamefont
  {Y.}~\bibnamefont {Yang}}, \bibinfo {author} {\bibfnamefont {U.}~\bibnamefont
  {Hohenester}}, \bibinfo {author} {\bibfnamefont {L.}~\bibnamefont {Huber}},
  \bibinfo {author} {\bibfnamefont {M.}~\bibnamefont {Wubs}}, \bibinfo {author}
  {\bibfnamefont {S.~D.}\ \bibnamefont {Liberato}}, \bibinfo {author}
  {\bibfnamefont {P.~A.~D.}\ \bibnamefont {Gon\c{c}alves}}, \bibinfo {author}
  {\bibfnamefont {F.~J.}\ \bibnamefont {{Garc\'{\i}a de Abajo}}}, \bibinfo
  {author} {\bibfnamefont {O.}~\bibnamefont {Hess}}, \bibinfo {author}
  {\bibfnamefont {I.}~\bibnamefont {Tarasenko}}, \bibinfo {author}
  {\bibfnamefont {J.~D.}\ \bibnamefont {Cox}}, \bibinfo {author} {\bibfnamefont
  {L.}~\bibnamefont {Jelver}}, \bibinfo {author} {\bibfnamefont {E.~J.~C.}\
  \bibnamefont {Dias}}, \bibinfo {author} {\bibfnamefont {M.~S.}\ \bibnamefont
  {S\'{a}nchez}}, \bibinfo {author} {\bibfnamefont {D.}~\bibnamefont
  {Margetis}}, \bibinfo {author} {\bibfnamefont {G.}~\bibnamefont
  {G\'{o}mez-Santos}}, \bibinfo {author} {\bibfnamefont {I.~M.}\ \bibnamefont
  {Vasilevskiy}}, \bibinfo {author} {\bibfnamefont {T.}~\bibnamefont
  {Stauber}}, \bibinfo {author} {\bibfnamefont {S.}~\bibnamefont {Tretyakov}},
  \bibinfo {author} {\bibfnamefont {C.}~\bibnamefont {Simovski}}, \bibinfo
  {author} {\bibfnamefont {S.}~\bibnamefont {Pakniyat}}, \bibinfo {author}
  {\bibfnamefont {J.~S.}\ \bibnamefont {G\'{o}mez-D\'{i}az}}, \bibinfo {author}
  {\bibfnamefont {I.~V.}\ \bibnamefont {Bondarev}}, \bibinfo {author}
  {\bibfnamefont {S.-A.}\ \bibnamefont {Biehs}}, \bibinfo {author}
  {\bibfnamefont {A.}~\bibnamefont {Boltasseva}}, \bibinfo {author}
  {\bibfnamefont {V.~M.}\ \bibnamefont {Shalaev}}, \bibinfo {author}
  {\bibfnamefont {A.~V.}\ \bibnamefont {Krasavin}}, \bibinfo {author}
  {\bibfnamefont {A.~V.}\ \bibnamefont {Zayats}}, \bibinfo {author}
  {\bibfnamefont {A.}~\bibnamefont {Al\`{u}}}, \bibinfo {author} {\bibfnamefont
  {J.-H.}\ \bibnamefont {Song}}, \bibinfo {author} {\bibfnamefont {M.~L.}\
  \bibnamefont {Brongersma}}, \bibinfo {author} {\bibfnamefont
  {U.}~\bibnamefont {Levy}}, \bibinfo {author} {\bibfnamefont {O.~Y.}\
  \bibnamefont {Long}}, \bibinfo {author} {\bibfnamefont {C.}~\bibnamefont
  {Guo}}, \bibinfo {author} {\bibfnamefont {S.}~\bibnamefont {Fan}}, \bibinfo
  {author} {\bibfnamefont {S.~I.}\ \bibnamefont {Bozhevolnyi}}, \bibinfo
  {author} {\bibfnamefont {A.}~\bibnamefont {Overvig}}, \bibinfo {author}
  {\bibfnamefont {F.~R.}\ \bibnamefont {Prud\^{e}ncio}}, \bibinfo {author}
  {\bibfnamefont {M.~G.}\ \bibnamefont {Silveirinha}}, \bibinfo {author}
  {\bibfnamefont {S.~A.~H.}\ \bibnamefont {Gangaraj}}, \bibinfo {author}
  {\bibfnamefont {C.}~\bibnamefont {Argyropoulos}}, \bibinfo {author}
  {\bibfnamefont {P.~A.}\ \bibnamefont {Huidobro}}, \bibinfo {author}
  {\bibfnamefont {E.}~\bibnamefont {Galiffi}}, \bibinfo {author} {\bibfnamefont
  {F.}~\bibnamefont {Yang}}, \bibinfo {author} {\bibfnamefont {J.~B.}\
  \bibnamefont {Pendry}},\ and\ \bibinfo {author} {\bibfnamefont {D.~A.~B.}\
  \bibnamefont {Miller}},\ }\href {https://doi.org/10.1364/OME.559374}
  {\bibfield  {journal} {\bibinfo  {journal} {Opt.\ Mater.\ Express}\ }\textbf
  {\bibinfo {volume} {15}},\ \bibinfo {pages} {1544} (\bibinfo {year}
  {2025})}\BibitemShut {NoStop}%
\bibitem [{\citenamefont {Gon{\c{c}}alves}\ and\ \citenamefont
  {Peres}(2016)}]{gonccalves2016introduction}%
  \BibitemOpen
  \bibfield  {author} {\bibinfo {author} {\bibfnamefont {P.~A.~D.}\
  \bibnamefont {Gon{\c{c}}alves}}\ and\ \bibinfo {author} {\bibfnamefont
  {N.~M.~R.}\ \bibnamefont {Peres}},\ }\href@noop {} {\emph {\bibinfo {title}
  {An introduction to graphene plasmonics}}}\ (\bibinfo  {publisher} {World
  Scientific},\ \bibinfo {year} {2016})\BibitemShut {NoStop}%
\bibitem [{\citenamefont {Murali}\ \emph {et~al.}(2009)\citenamefont {Murali},
  \citenamefont {Yang}, \citenamefont {Brenner}, \citenamefont {Beck},\ and\
  \citenamefont {Meindl}}]{murali2009breakdown}%
  \BibitemOpen
  \bibfield  {author} {\bibinfo {author} {\bibfnamefont {R.}~\bibnamefont
  {Murali}}, \bibinfo {author} {\bibfnamefont {Y.}~\bibnamefont {Yang}},
  \bibinfo {author} {\bibfnamefont {K.}~\bibnamefont {Brenner}}, \bibinfo
  {author} {\bibfnamefont {T.}~\bibnamefont {Beck}},\ and\ \bibinfo {author}
  {\bibfnamefont {J.~D.}\ \bibnamefont {Meindl}},\ }\href
  {https://doi.org/10.1063/1.3147183} {\bibfield  {journal} {\bibinfo
  {journal} {Appl.\ Phys.\ Lett.}\ }\textbf {\bibinfo {volume} {94}},\ \bibinfo
  {pages} {243114} (\bibinfo {year} {2009})}\BibitemShut {NoStop}%
\bibitem [{\citenamefont {Liao}\ \emph {et~al.}(2010)\citenamefont {Liao},
  \citenamefont {Lin}, \citenamefont {Bao}, \citenamefont {Cheng},
  \citenamefont {Bai}, \citenamefont {Liu}, \citenamefont {Qu}, \citenamefont
  {Wang}, \citenamefont {Huang},\ and\ \citenamefont {Duan}}]{liao2010high}%
  \BibitemOpen
  \bibfield  {author} {\bibinfo {author} {\bibfnamefont {L.}~\bibnamefont
  {Liao}}, \bibinfo {author} {\bibfnamefont {Y.-C.}\ \bibnamefont {Lin}},
  \bibinfo {author} {\bibfnamefont {M.}~\bibnamefont {Bao}}, \bibinfo {author}
  {\bibfnamefont {R.}~\bibnamefont {Cheng}}, \bibinfo {author} {\bibfnamefont
  {J.}~\bibnamefont {Bai}}, \bibinfo {author} {\bibfnamefont {Y.}~\bibnamefont
  {Liu}}, \bibinfo {author} {\bibfnamefont {Y.}~\bibnamefont {Qu}}, \bibinfo
  {author} {\bibfnamefont {K.~L.}\ \bibnamefont {Wang}}, \bibinfo {author}
  {\bibfnamefont {Y.}~\bibnamefont {Huang}},\ and\ \bibinfo {author}
  {\bibfnamefont {X.}~\bibnamefont {Duan}},\ }\href
  {https://doi.org/10.1038/nature09405} {\bibfield  {journal} {\bibinfo
  {journal} {Nature}\ }\textbf {\bibinfo {volume} {467}},\ \bibinfo {pages}
  {305} (\bibinfo {year} {2010})}\BibitemShut {NoStop}%
\bibitem [{\citenamefont {Son}\ \emph {et~al.}(2017)\citenamefont {Son},
  \citenamefont {{\v{S}}i{\v{s}}kins}, \citenamefont {Mullan}, \citenamefont
  {Yin}, \citenamefont {Kravets}, \citenamefont {Kozikov}, \citenamefont
  {Ozdemir}, \citenamefont {Alhazmi}, \citenamefont {Holwill}, \citenamefont
  {Watanabe}, \citenamefont {Taniguchi}, \citenamefont {Ghazaryan},
  \citenamefont {Novoselov}, \citenamefont {Fal'ko},\ and\ \citenamefont
  {Mishchenko}}]{son2017graphene}%
  \BibitemOpen
  \bibfield  {author} {\bibinfo {author} {\bibfnamefont {S.-K.}\ \bibnamefont
  {Son}}, \bibinfo {author} {\bibfnamefont {M.}~\bibnamefont
  {{\v{S}}i{\v{s}}kins}}, \bibinfo {author} {\bibfnamefont {C.}~\bibnamefont
  {Mullan}}, \bibinfo {author} {\bibfnamefont {J.}~\bibnamefont {Yin}},
  \bibinfo {author} {\bibfnamefont {V.~G.}\ \bibnamefont {Kravets}}, \bibinfo
  {author} {\bibfnamefont {A.}~\bibnamefont {Kozikov}}, \bibinfo {author}
  {\bibfnamefont {S.}~\bibnamefont {Ozdemir}}, \bibinfo {author} {\bibfnamefont
  {M.}~\bibnamefont {Alhazmi}}, \bibinfo {author} {\bibfnamefont
  {M.}~\bibnamefont {Holwill}}, \bibinfo {author} {\bibfnamefont
  {K.}~\bibnamefont {Watanabe}}, \bibinfo {author} {\bibfnamefont
  {T.}~\bibnamefont {Taniguchi}}, \bibinfo {author} {\bibfnamefont
  {D.}~\bibnamefont {Ghazaryan}}, \bibinfo {author} {\bibfnamefont {K.~S.}\
  \bibnamefont {Novoselov}}, \bibinfo {author} {\bibfnamefont {V.~I.}\
  \bibnamefont {Fal'ko}},\ and\ \bibinfo {author} {\bibfnamefont
  {A.}~\bibnamefont {Mishchenko}},\ }\href
  {https://doi.org/10.1088/2053-1583/aa97b5} {\bibfield  {journal} {\bibinfo
  {journal} {2D Mater.}\ }\textbf {\bibinfo {volume} {5}},\ \bibinfo {pages}
  {011006} (\bibinfo {year} {2017})}\BibitemShut {NoStop}%
\bibitem [{\citenamefont {Barajas‑Aguilar}\ \emph {et~al.}(2024)\citenamefont
  {Barajas‑Aguilar}, \citenamefont {Zion}, \citenamefont {Sequeira},
  \citenamefont {Barabas}, \citenamefont {Taniguchi}, \citenamefont {Watanabe},
  \citenamefont {Barrett}, \citenamefont {Scaffidi},\ and\ \citenamefont
  {Sanchez‑Yamagishi}}]{barajas2024electrically}%
  \BibitemOpen
  \bibfield  {author} {\bibinfo {author} {\bibfnamefont {A.~H.}\ \bibnamefont
  {Barajas‑Aguilar}}, \bibinfo {author} {\bibfnamefont {J.}~\bibnamefont
  {Zion}}, \bibinfo {author} {\bibfnamefont {I.}~\bibnamefont {Sequeira}},
  \bibinfo {author} {\bibfnamefont {A.~Z.}\ \bibnamefont {Barabas}}, \bibinfo
  {author} {\bibfnamefont {T.}~\bibnamefont {Taniguchi}}, \bibinfo {author}
  {\bibfnamefont {K.}~\bibnamefont {Watanabe}}, \bibinfo {author}
  {\bibfnamefont {E.~B.}\ \bibnamefont {Barrett}}, \bibinfo {author}
  {\bibfnamefont {T.}~\bibnamefont {Scaffidi}},\ and\ \bibinfo {author}
  {\bibfnamefont {J.~D.}\ \bibnamefont {Sanchez‑Yamagishi}},\ }\href
  {https://doi.org/10.1038/s41467-024-46819-2} {\bibfield  {journal} {\bibinfo
  {journal} {Nat.\ Commun.}\ }\textbf {\bibinfo {volume} {15}},\ \bibinfo
  {pages} {2550} (\bibinfo {year} {2024})}\BibitemShut {NoStop}%
\bibitem [{\citenamefont {Sabbaghi}\ \emph {et~al.}(2015)\citenamefont
  {Sabbaghi}, \citenamefont {Lee}, \citenamefont {Stauber},\ and\ \citenamefont
  {Kim}}]{sabbaghi2015drift}%
  \BibitemOpen
  \bibfield  {author} {\bibinfo {author} {\bibfnamefont {M.}~\bibnamefont
  {Sabbaghi}}, \bibinfo {author} {\bibfnamefont {H.-W.}\ \bibnamefont {Lee}},
  \bibinfo {author} {\bibfnamefont {T.}~\bibnamefont {Stauber}},\ and\ \bibinfo
  {author} {\bibfnamefont {K.~S.}\ \bibnamefont {Kim}},\ }\href
  {https://doi.org/10.1103/PhysRevB.92.195429} {\bibfield  {journal} {\bibinfo
  {journal} {Phys.\ Rev.\ B}\ }\textbf {\bibinfo {volume} {92}},\ \bibinfo
  {pages} {195429} (\bibinfo {year} {2015})}\BibitemShut {NoStop}%
\bibitem [{\citenamefont {Duppen}\ \emph {et~al.}(2016)\citenamefont {Duppen},
  \citenamefont {Tomadin}, \citenamefont {Grigorenko},\ and\ \citenamefont
  {Polini}}]{vanduppen2016current}%
  \BibitemOpen
  \bibfield  {author} {\bibinfo {author} {\bibfnamefont {B.~V.}\ \bibnamefont
  {Duppen}}, \bibinfo {author} {\bibfnamefont {A.}~\bibnamefont {Tomadin}},
  \bibinfo {author} {\bibfnamefont {A.~N.}\ \bibnamefont {Grigorenko}},\ and\
  \bibinfo {author} {\bibfnamefont {M.}~\bibnamefont {Polini}},\ }\href
  {https://doi.org/10.1088/2053-1583/3/1/015011} {\bibfield  {journal}
  {\bibinfo  {journal} {2D Mater.}\ }\textbf {\bibinfo {volume} {3}},\ \bibinfo
  {pages} {015011} (\bibinfo {year} {2016})}\BibitemShut {NoStop}%
\bibitem [{\citenamefont {Wenger}\ \emph {et~al.}(2018)\citenamefont {Wenger},
  \citenamefont {Viola}, \citenamefont {Kinaret}, \citenamefont
  {Fogelstr{\"o}m},\ and\ \citenamefont {Tassin}}]{wenger2018current}%
  \BibitemOpen
  \bibfield  {author} {\bibinfo {author} {\bibfnamefont {T.}~\bibnamefont
  {Wenger}}, \bibinfo {author} {\bibfnamefont {G.}~\bibnamefont {Viola}},
  \bibinfo {author} {\bibfnamefont {J.}~\bibnamefont {Kinaret}}, \bibinfo
  {author} {\bibfnamefont {M.}~\bibnamefont {Fogelstr{\"o}m}},\ and\ \bibinfo
  {author} {\bibfnamefont {P.}~\bibnamefont {Tassin}},\ }\href
  {https://doi.org/10.1103/PhysRevB.97.085419} {\bibfield  {journal} {\bibinfo
  {journal} {Phys.\ Rev.\ B}\ }\textbf {\bibinfo {volume} {97}},\ \bibinfo
  {pages} {085419} (\bibinfo {year} {2018})}\BibitemShut {NoStop}%
\bibitem [{\citenamefont {Dong}\ \emph {et~al.}(2021)\citenamefont {Dong},
  \citenamefont {Xiong}, \citenamefont {Phinney}, \citenamefont {Sun},
  \citenamefont {Jing}, \citenamefont {McLeod}, \citenamefont {Zhang},
  \citenamefont {Liu}, \citenamefont {Ruta}, \citenamefont {Gao}, \citenamefont
  {Dong}, \citenamefont {Pan}, \citenamefont {Edgar}, \citenamefont
  {Jarillo-Herrero}, \citenamefont {Levitov}, \citenamefont {Millis},
  \citenamefont {Fogler}, \citenamefont {Bandurin},\ and\ \citenamefont
  {Basov}}]{dong2021fizeau}%
  \BibitemOpen
  \bibfield  {author} {\bibinfo {author} {\bibfnamefont {Y.}~\bibnamefont
  {Dong}}, \bibinfo {author} {\bibfnamefont {L.}~\bibnamefont {Xiong}},
  \bibinfo {author} {\bibfnamefont {I.~Y.}\ \bibnamefont {Phinney}}, \bibinfo
  {author} {\bibfnamefont {Z.}~\bibnamefont {Sun}}, \bibinfo {author}
  {\bibfnamefont {R.}~\bibnamefont {Jing}}, \bibinfo {author} {\bibfnamefont
  {A.~S.}\ \bibnamefont {McLeod}}, \bibinfo {author} {\bibfnamefont
  {S.}~\bibnamefont {Zhang}}, \bibinfo {author} {\bibfnamefont
  {S.}~\bibnamefont {Liu}}, \bibinfo {author} {\bibfnamefont {F.~L.}\
  \bibnamefont {Ruta}}, \bibinfo {author} {\bibfnamefont {H.}~\bibnamefont
  {Gao}}, \bibinfo {author} {\bibfnamefont {Z.}~\bibnamefont {Dong}}, \bibinfo
  {author} {\bibfnamefont {R.}~\bibnamefont {Pan}}, \bibinfo {author}
  {\bibfnamefont {J.~H.}\ \bibnamefont {Edgar}}, \bibinfo {author}
  {\bibfnamefont {P.}~\bibnamefont {Jarillo-Herrero}}, \bibinfo {author}
  {\bibfnamefont {L.~S.}\ \bibnamefont {Levitov}}, \bibinfo {author}
  {\bibfnamefont {A.~J.}\ \bibnamefont {Millis}}, \bibinfo {author}
  {\bibfnamefont {M.~M.}\ \bibnamefont {Fogler}}, \bibinfo {author}
  {\bibfnamefont {D.~A.}\ \bibnamefont {Bandurin}},\ and\ \bibinfo {author}
  {\bibfnamefont {D.~N.}\ \bibnamefont {Basov}},\ }\href
  {https://doi.org/10.1038/s41586-021-03640-x} {\bibfield  {journal} {\bibinfo
  {journal} {Nature}\ }\textbf {\bibinfo {volume} {594}},\ \bibinfo {pages}
  {513} (\bibinfo {year} {2021})}\BibitemShut {NoStop}%
\bibitem [{\citenamefont {Zhao}\ \emph {et~al.}(2021)\citenamefont {Zhao},
  \citenamefont {Zhao}, \citenamefont {Li}, \citenamefont {Wang}, \citenamefont
  {Wang}, \citenamefont {{Iqbal Bakti Utama}}, \citenamefont {Kahn},
  \citenamefont {Jiang}, \citenamefont {Xiao}, \citenamefont {Yoo},
  \citenamefont {Watanabe}, \citenamefont {Taniguchi}, \citenamefont {Zettl},\
  and\ \citenamefont {Wang}}]{zhao2021efficient}%
  \BibitemOpen
  \bibfield  {author} {\bibinfo {author} {\bibfnamefont {W.}~\bibnamefont
  {Zhao}}, \bibinfo {author} {\bibfnamefont {S.}~\bibnamefont {Zhao}}, \bibinfo
  {author} {\bibfnamefont {H.}~\bibnamefont {Li}}, \bibinfo {author}
  {\bibfnamefont {S.}~\bibnamefont {Wang}}, \bibinfo {author} {\bibfnamefont
  {S.}~\bibnamefont {Wang}}, \bibinfo {author} {\bibfnamefont {M.}~\bibnamefont
  {{Iqbal Bakti Utama}}}, \bibinfo {author} {\bibfnamefont {S.}~\bibnamefont
  {Kahn}}, \bibinfo {author} {\bibfnamefont {Y.}~\bibnamefont {Jiang}},
  \bibinfo {author} {\bibfnamefont {X.}~\bibnamefont {Xiao}}, \bibinfo {author}
  {\bibfnamefont {S.}~\bibnamefont {Yoo}}, \bibinfo {author} {\bibfnamefont
  {K.}~\bibnamefont {Watanabe}}, \bibinfo {author} {\bibfnamefont
  {T.}~\bibnamefont {Taniguchi}}, \bibinfo {author} {\bibfnamefont
  {A.}~\bibnamefont {Zettl}},\ and\ \bibinfo {author} {\bibfnamefont
  {F.}~\bibnamefont {Wang}},\ }\href
  {https://doi.org/10.1038/s41586-021-03574-4} {\bibfield  {journal} {\bibinfo
  {journal} {Nature}\ }\textbf {\bibinfo {volume} {594}},\ \bibinfo {pages}
  {517} (\bibinfo {year} {2021})}\BibitemShut {NoStop}%
\bibitem [{\citenamefont {Dong}\ \emph {et~al.}(2025)\citenamefont {Dong},
  \citenamefont {Sun}, \citenamefont {Phinney}, \citenamefont {Sun},
  \citenamefont {Andersen}, \citenamefont {Xiong}, \citenamefont {Shao},
  \citenamefont {Zhang}, \citenamefont {Rikhter}, \citenamefont {Liu},
  \citenamefont {Jarillo‑Herrero}, \citenamefont {Kim}, \citenamefont {Dean},
  \citenamefont {Millis}, \citenamefont {Fogler}, \citenamefont {Bandurin},\
  and\ \citenamefont {Basov}}]{dong2025current}%
  \BibitemOpen
  \bibfield  {author} {\bibinfo {author} {\bibfnamefont {Y.}~\bibnamefont
  {Dong}}, \bibinfo {author} {\bibfnamefont {Z.}~\bibnamefont {Sun}}, \bibinfo
  {author} {\bibfnamefont {I.~Y.}\ \bibnamefont {Phinney}}, \bibinfo {author}
  {\bibfnamefont {D.}~\bibnamefont {Sun}}, \bibinfo {author} {\bibfnamefont
  {T.~I.}\ \bibnamefont {Andersen}}, \bibinfo {author} {\bibfnamefont
  {L.}~\bibnamefont {Xiong}}, \bibinfo {author} {\bibfnamefont
  {Y.}~\bibnamefont {Shao}}, \bibinfo {author} {\bibfnamefont {S.}~\bibnamefont
  {Zhang}}, \bibinfo {author} {\bibfnamefont {A.}~\bibnamefont {Rikhter}},
  \bibinfo {author} {\bibfnamefont {S.}~\bibnamefont {Liu}}, \bibinfo {author}
  {\bibfnamefont {P.}~\bibnamefont {Jarillo‑Herrero}}, \bibinfo {author}
  {\bibfnamefont {P.}~\bibnamefont {Kim}}, \bibinfo {author} {\bibfnamefont
  {C.~R.}\ \bibnamefont {Dean}}, \bibinfo {author} {\bibfnamefont {A.~J.}\
  \bibnamefont {Millis}}, \bibinfo {author} {\bibfnamefont {M.~M.}\
  \bibnamefont {Fogler}}, \bibinfo {author} {\bibfnamefont {D.~A.}\
  \bibnamefont {Bandurin}},\ and\ \bibinfo {author} {\bibfnamefont {D.~N.}\
  \bibnamefont {Basov}},\ }\href {https://doi.org/10.1038/s41467-025-58953-6}
  {\bibfield  {journal} {\bibinfo  {journal} {Nat.\ Commun.}\ }\textbf
  {\bibinfo {volume} {16}},\ \bibinfo {pages} {3861} (\bibinfo {year}
  {2025})}\BibitemShut {NoStop}%
\bibitem [{\citenamefont {Blevins}\ and\ \citenamefont
  {Boriskina}(2024)}]{blevins2024plasmon}%
  \BibitemOpen
  \bibfield  {author} {\bibinfo {author} {\bibfnamefont {M.~G.}\ \bibnamefont
  {Blevins}}\ and\ \bibinfo {author} {\bibfnamefont {S.~V.}\ \bibnamefont
  {Boriskina}},\ }\href {https://doi.org/10.1021/acsphotonics.3c01416}
  {\bibfield  {journal} {\bibinfo  {journal} {ACS Photonics}\ }\textbf
  {\bibinfo {volume} {11}},\ \bibinfo {pages} {537} (\bibinfo {year}
  {2024})}\BibitemShut {NoStop}%
\bibitem [{\citenamefont {Morgado}\ and\ \citenamefont
  {Silveirinha}(2017)}]{morgado2017negative}%
  \BibitemOpen
  \bibfield  {author} {\bibinfo {author} {\bibfnamefont {T.~A.}\ \bibnamefont
  {Morgado}}\ and\ \bibinfo {author} {\bibfnamefont {M.~G.}\ \bibnamefont
  {Silveirinha}},\ }\href {https://doi.org/10.1103/PhysRevLett.119.133901}
  {\bibfield  {journal} {\bibinfo  {journal} {Phys.\ Rev.\ Lett.}\ }\textbf
  {\bibinfo {volume} {119}},\ \bibinfo {pages} {133901} (\bibinfo {year}
  {2017})}\BibitemShut {NoStop}%
\bibitem [{\citenamefont {Morgado}\ and\ \citenamefont
  {Silveirinha}(2018)}]{morgado2018drift}%
  \BibitemOpen
  \bibfield  {author} {\bibinfo {author} {\bibfnamefont {T.~A.}\ \bibnamefont
  {Morgado}}\ and\ \bibinfo {author} {\bibfnamefont {M.~G.}\ \bibnamefont
  {Silveirinha}},\ }\href {https://doi.org/10.1021/acsphotonics.8b00987}
  {\bibfield  {journal} {\bibinfo  {journal} {ACS Photonics}\ }\textbf
  {\bibinfo {volume} {5}},\ \bibinfo {pages} {4253} (\bibinfo {year}
  {2018})}\BibitemShut {NoStop}%
\bibitem [{\citenamefont {Morgado}\ and\ \citenamefont
  {Silveirinha}(2020)}]{morgado2020nonlocal}%
  \BibitemOpen
  \bibfield  {author} {\bibinfo {author} {\bibfnamefont {T.~A.}\ \bibnamefont
  {Morgado}}\ and\ \bibinfo {author} {\bibfnamefont {M.~G.}\ \bibnamefont
  {Silveirinha}},\ }\href {https://doi.org/10.1103/PhysRevB.102.075102}
  {\bibfield  {journal} {\bibinfo  {journal} {Phys.\ Rev.\ B}\ }\textbf
  {\bibinfo {volume} {102}},\ \bibinfo {pages} {075102} (\bibinfo {year}
  {2020})}\BibitemShut {NoStop}%
\bibitem [{\citenamefont {Correas-Serrano}\ and\ \citenamefont
  {Gomez-Diaz}(2019)}]{correas2019nonreciprocal}%
  \BibitemOpen
  \bibfield  {author} {\bibinfo {author} {\bibfnamefont {D.}~\bibnamefont
  {Correas-Serrano}}\ and\ \bibinfo {author} {\bibfnamefont {J.~S.}\
  \bibnamefont {Gomez-Diaz}},\ }\href
  {https://doi.org/10.1103/PhysRevB.100.081410} {\bibfield  {journal} {\bibinfo
   {journal} {Phys.\ Rev.\ B}\ }\textbf {\bibinfo {volume} {100}},\ \bibinfo
  {pages} {081410} (\bibinfo {year} {2019})}\BibitemShut {NoStop}%
\bibitem [{\citenamefont {Prud{\^e}ncio}\ and\ \citenamefont
  {Silveirinha}(2021)}]{prudencio2021asymmetric}%
  \BibitemOpen
  \bibfield  {author} {\bibinfo {author} {\bibfnamefont {F.~R.}\ \bibnamefont
  {Prud{\^e}ncio}}\ and\ \bibinfo {author} {\bibfnamefont {M.~G.}\ \bibnamefont
  {Silveirinha}},\ }\href {https://doi.org/10.1007/s11468-020-01215-6}
  {\bibfield  {journal} {\bibinfo  {journal} {Plasmonics}\ }\textbf {\bibinfo
  {volume} {16}},\ \bibinfo {pages} {19} (\bibinfo {year} {2021})}\BibitemShut
  {NoStop}%
\bibitem [{\citenamefont {Morgado}\ and\ \citenamefont
  {Silveirinha}(2022)}]{morgado2022directional}%
  \BibitemOpen
  \bibfield  {author} {\bibinfo {author} {\bibfnamefont {T.~A.}\ \bibnamefont
  {Morgado}}\ and\ \bibinfo {author} {\bibfnamefont {M.~G.}\ \bibnamefont
  {Silveirinha}},\ }\href {https://doi.org/10.1515/nanoph-2022-0380} {\bibfield
   {journal} {\bibinfo  {journal} {Nanophotonics}\ }\textbf {\bibinfo {volume}
  {11}},\ \bibinfo {pages} {4929} (\bibinfo {year} {2022})}\BibitemShut
  {NoStop}%
\bibitem [{\citenamefont {Christensen}\ \emph {et~al.}(2012)\citenamefont
  {Christensen}, \citenamefont {Manjavacas}, \citenamefont {Thongrattanasiri},
  \citenamefont {Koppens},\ and\ \citenamefont {{Garc\'{\i}a de
  Abajo}}}]{christensen2012graphene}%
  \BibitemOpen
  \bibfield  {author} {\bibinfo {author} {\bibfnamefont {J.}~\bibnamefont
  {Christensen}}, \bibinfo {author} {\bibfnamefont {A.}~\bibnamefont
  {Manjavacas}}, \bibinfo {author} {\bibfnamefont {S.}~\bibnamefont
  {Thongrattanasiri}}, \bibinfo {author} {\bibfnamefont {F.~H.~L.}\
  \bibnamefont {Koppens}},\ and\ \bibinfo {author} {\bibfnamefont {F.~J.}\
  \bibnamefont {{Garc\'{\i}a de Abajo}}},\ }\href
  {https://doi.org/10.1021/nn2037626} {\bibfield  {journal} {\bibinfo
  {journal} {ACS Nano}\ }\textbf {\bibinfo {volume} {6}},\ \bibinfo {pages}
  {431} (\bibinfo {year} {2012})}\BibitemShut {NoStop}%
\bibitem [{\citenamefont {Huidobro}\ \emph {et~al.}(2012)\citenamefont
  {Huidobro}, \citenamefont {Nikitin}, \citenamefont {Gonz\'{a}lez-Ballestero},
  \citenamefont {Mart\'{i}n-Moreno},\ and\ \citenamefont
  {Garc\'{i}a-Vidal}}]{huidobro2012superradiance}%
  \BibitemOpen
  \bibfield  {author} {\bibinfo {author} {\bibfnamefont {P.~A.}\ \bibnamefont
  {Huidobro}}, \bibinfo {author} {\bibfnamefont {A.~Y.}\ \bibnamefont
  {Nikitin}}, \bibinfo {author} {\bibfnamefont {C.}~\bibnamefont
  {Gonz\'{a}lez-Ballestero}}, \bibinfo {author} {\bibfnamefont
  {L.}~\bibnamefont {Mart\'{i}n-Moreno}},\ and\ \bibinfo {author}
  {\bibfnamefont {F.~J.}\ \bibnamefont {Garc\'{i}a-Vidal}},\ }\href
  {https://doi.org/10.1103/PhysRevB.85.155438} {\bibfield  {journal} {\bibinfo
  {journal} {Phys.\ Rev.\ B}\ }\textbf {\bibinfo {volume} {85}},\ \bibinfo
  {pages} {155438} (\bibinfo {year} {2012})}\BibitemShut {NoStop}%
\bibitem [{\citenamefont {Mart{\'\i}n-Moreno}\ \emph
  {et~al.}(2015)\citenamefont {Mart{\'\i}n-Moreno}, \citenamefont {{Garc\'{\i}a
  de Abajo}},\ and\ \citenamefont
  {Garc{\'\i}a-Vidal}}]{martinmoreno2015ultraefficient}%
  \BibitemOpen
  \bibfield  {author} {\bibinfo {author} {\bibfnamefont {L.}~\bibnamefont
  {Mart{\'\i}n-Moreno}}, \bibinfo {author} {\bibfnamefont {F.~J.}\ \bibnamefont
  {{Garc\'{\i}a de Abajo}}},\ and\ \bibinfo {author} {\bibfnamefont {F.~J.}\
  \bibnamefont {Garc{\'\i}a-Vidal}},\ }\href
  {https://doi.org/10.1103/PhysRevLett.115.173601} {\bibfield  {journal}
  {\bibinfo  {journal} {Phys.\ Rev.\ Lett.}\ }\textbf {\bibinfo {volume}
  {115}},\ \bibinfo {pages} {173601} (\bibinfo {year} {2015})}\BibitemShut
  {NoStop}%
\bibitem [{\citenamefont {Lamata}\ \emph {et~al.}(2015)\citenamefont {Lamata},
  \citenamefont {Alonso-Gonz{\'a}lez}, \citenamefont {Hillenbrand},\ and\
  \citenamefont {Nikitin}}]{soto2015plasmons}%
  \BibitemOpen
  \bibfield  {author} {\bibinfo {author} {\bibfnamefont {I.~S.}\ \bibnamefont
  {Lamata}}, \bibinfo {author} {\bibfnamefont {P.}~\bibnamefont
  {Alonso-Gonz{\'a}lez}}, \bibinfo {author} {\bibfnamefont {R.}~\bibnamefont
  {Hillenbrand}},\ and\ \bibinfo {author} {\bibfnamefont {A.~Y.}\ \bibnamefont
  {Nikitin}},\ }\href {https://doi.org/10.1021/ph500377u} {\bibfield  {journal}
  {\bibinfo  {journal} {ACS Photonics}\ }\textbf {\bibinfo {volume} {2}},\
  \bibinfo {pages} {280} (\bibinfo {year} {2015})}\BibitemShut {NoStop}%
\bibitem [{\citenamefont {{de Vega}}\ \emph {et~al.}(2016)\citenamefont {{de
  Vega}}, \citenamefont {Cox},\ and\ \citenamefont {{Garc\'{\i}a de
  Abajo}}}]{devega2016plasmons}%
  \BibitemOpen
  \bibfield  {author} {\bibinfo {author} {\bibfnamefont {S.}~\bibnamefont {{de
  Vega}}}, \bibinfo {author} {\bibfnamefont {J.~D.}\ \bibnamefont {Cox}},\ and\
  \bibinfo {author} {\bibfnamefont {F.~J.}\ \bibnamefont {{Garc\'{\i}a de
  Abajo}}},\ }\href {https://doi.org/10.1103/PhysRevB.94.075447} {\bibfield
  {journal} {\bibinfo  {journal} {Phys.\ Rev.\ B}\ }\textbf {\bibinfo {volume}
  {94}},\ \bibinfo {pages} {075447} (\bibinfo {year} {2016})}\BibitemShut
  {NoStop}%
\bibitem [{\citenamefont {Calaj{\'o}}\ \emph {et~al.}(2023)\citenamefont
  {Calaj{\'o}}, \citenamefont {Jenke}, \citenamefont {Rozema}, \citenamefont
  {Walther}, \citenamefont {Chang},\ and\ \citenamefont
  {Cox}}]{calajo2023nonlinear}%
  \BibitemOpen
  \bibfield  {author} {\bibinfo {author} {\bibfnamefont {G.}~\bibnamefont
  {Calaj{\'o}}}, \bibinfo {author} {\bibfnamefont {P.~K.}\ \bibnamefont
  {Jenke}}, \bibinfo {author} {\bibfnamefont {L.~A.}\ \bibnamefont {Rozema}},
  \bibinfo {author} {\bibfnamefont {P.}~\bibnamefont {Walther}}, \bibinfo
  {author} {\bibfnamefont {D.~E.}\ \bibnamefont {Chang}},\ and\ \bibinfo
  {author} {\bibfnamefont {J.~D.}\ \bibnamefont {Cox}},\ }\href
  {https://doi.org/10.1103/PhysRevResearch.5.013188} {\bibfield  {journal}
  {\bibinfo  {journal} {Phys.\ Rev.\ Research}\ }\textbf {\bibinfo {volume}
  {5}},\ \bibinfo {pages} {013188} (\bibinfo {year} {2023})}\BibitemShut
  {NoStop}%
\bibitem [{\citenamefont {Baringhaus}\ \emph {et~al.}(2014)\citenamefont
  {Baringhaus}, \citenamefont {Ruan}, \citenamefont {Edler}, \citenamefont
  {Tejeda}, \citenamefont {Sicot}, \citenamefont {Taleb‑Ibrahimi},
  \citenamefont {Li}, \citenamefont {Jiang}, \citenamefont {Conrad},
  \citenamefont {Berger}, \citenamefont {Tegenkamp},\ and\ \citenamefont
  {de~Heer}}]{baringhaus2014exceptional}%
  \BibitemOpen
  \bibfield  {author} {\bibinfo {author} {\bibfnamefont {J.}~\bibnamefont
  {Baringhaus}}, \bibinfo {author} {\bibfnamefont {M.}~\bibnamefont {Ruan}},
  \bibinfo {author} {\bibfnamefont {F.}~\bibnamefont {Edler}}, \bibinfo
  {author} {\bibfnamefont {A.}~\bibnamefont {Tejeda}}, \bibinfo {author}
  {\bibfnamefont {M.}~\bibnamefont {Sicot}}, \bibinfo {author} {\bibfnamefont
  {A.}~\bibnamefont {Taleb‑Ibrahimi}}, \bibinfo {author} {\bibfnamefont
  {A.}~\bibnamefont {Li}}, \bibinfo {author} {\bibfnamefont {Z.}~\bibnamefont
  {Jiang}}, \bibinfo {author} {\bibfnamefont {E.~H.}\ \bibnamefont {Conrad}},
  \bibinfo {author} {\bibfnamefont {C.}~\bibnamefont {Berger}}, \bibinfo
  {author} {\bibfnamefont {C.}~\bibnamefont {Tegenkamp}},\ and\ \bibinfo
  {author} {\bibfnamefont {W.~A.}\ \bibnamefont {de~Heer}},\ }\href
  {https://doi.org/10.1038/nature12952} {\bibfield  {journal} {\bibinfo
  {journal} {Nature}\ }\textbf {\bibinfo {volume} {506}},\ \bibinfo {pages}
  {349} (\bibinfo {year} {2014})}\BibitemShut {NoStop}%
\bibitem [{\citenamefont {Laird}\ \emph {et~al.}(2015)\citenamefont {Laird},
  \citenamefont {Kuemmeth}, \citenamefont {Steele}, \citenamefont
  {Grove-Rasmussen}, \citenamefont {Nyg{\aa}rd}, \citenamefont {Flensberg},\
  and\ \citenamefont {Kouwenhoven}}]{laird2015quantum}%
  \BibitemOpen
  \bibfield  {author} {\bibinfo {author} {\bibfnamefont {E.~A.}\ \bibnamefont
  {Laird}}, \bibinfo {author} {\bibfnamefont {F.}~\bibnamefont {Kuemmeth}},
  \bibinfo {author} {\bibfnamefont {G.~A.}\ \bibnamefont {Steele}}, \bibinfo
  {author} {\bibfnamefont {K.}~\bibnamefont {Grove-Rasmussen}}, \bibinfo
  {author} {\bibfnamefont {J.}~\bibnamefont {Nyg{\aa}rd}}, \bibinfo {author}
  {\bibfnamefont {K.}~\bibnamefont {Flensberg}},\ and\ \bibinfo {author}
  {\bibfnamefont {L.~P.}\ \bibnamefont {Kouwenhoven}},\ }\href
  {https://doi.org/10.1103/RevModPhys.87.703} {\bibfield  {journal} {\bibinfo
  {journal} {Rev.\ Mod.\ Phys.}\ }\textbf {\bibinfo {volume} {87}},\ \bibinfo
  {pages} {703} (\bibinfo {year} {2015})}\BibitemShut {NoStop}%
\bibitem [{\citenamefont {Wang}\ \emph {et~al.}(2021)\citenamefont {Wang},
  \citenamefont {Wang}, \citenamefont {Ma}, \citenamefont {Chen}, \citenamefont
  {Jiang}, \citenamefont {Chen}, \citenamefont {Xie}, \citenamefont {Li},\ and\
  \citenamefont {Wang}}]{wang2021graphene}%
  \BibitemOpen
  \bibfield  {author} {\bibinfo {author} {\bibfnamefont {H.}~\bibnamefont
  {Wang}}, \bibinfo {author} {\bibfnamefont {H.~S.}\ \bibnamefont {Wang}},
  \bibinfo {author} {\bibfnamefont {C.}~\bibnamefont {Ma}}, \bibinfo {author}
  {\bibfnamefont {L.}~\bibnamefont {Chen}}, \bibinfo {author} {\bibfnamefont
  {C.}~\bibnamefont {Jiang}}, \bibinfo {author} {\bibfnamefont
  {C.}~\bibnamefont {Chen}}, \bibinfo {author} {\bibfnamefont {X.}~\bibnamefont
  {Xie}}, \bibinfo {author} {\bibfnamefont {A.-P.}\ \bibnamefont {Li}},\ and\
  \bibinfo {author} {\bibfnamefont {X.}~\bibnamefont {Wang}},\ }\href
  {https://doi.org/10.1038/s42254-021-00370-x} {\bibfield  {journal} {\bibinfo
  {journal} {Nat.\ Rev.\ Phys.}\ }\textbf {\bibinfo {volume} {3}},\ \bibinfo
  {pages} {791} (\bibinfo {year} {2021})}\BibitemShut {NoStop}%
\bibitem [{\citenamefont {Yao}\ \emph {et~al.}(2023)\citenamefont {Yao},
  \citenamefont {Zhang}, \citenamefont {Kong}, \citenamefont {Hinaut},
  \citenamefont {Pawlak}, \citenamefont {Okuno}, \citenamefont {Graf},
  \citenamefont {Horton}, \citenamefont {Coles}, \citenamefont {Meyer},
  \citenamefont {Bogani}, \citenamefont {Bonn}, \citenamefont {Wang},
  \citenamefont {Müllen},\ and\ \citenamefont {Narita}}]{yao2023n}%
  \BibitemOpen
  \bibfield  {author} {\bibinfo {author} {\bibfnamefont {X.}~\bibnamefont
  {Yao}}, \bibinfo {author} {\bibfnamefont {H.}~\bibnamefont {Zhang}}, \bibinfo
  {author} {\bibfnamefont {F.}~\bibnamefont {Kong}}, \bibinfo {author}
  {\bibfnamefont {A.}~\bibnamefont {Hinaut}}, \bibinfo {author} {\bibfnamefont
  {R.}~\bibnamefont {Pawlak}}, \bibinfo {author} {\bibfnamefont
  {M.}~\bibnamefont {Okuno}}, \bibinfo {author} {\bibfnamefont
  {R.}~\bibnamefont {Graf}}, \bibinfo {author} {\bibfnamefont {P.~N.}\
  \bibnamefont {Horton}}, \bibinfo {author} {\bibfnamefont {S.~J.}\
  \bibnamefont {Coles}}, \bibinfo {author} {\bibfnamefont {E.}~\bibnamefont
  {Meyer}}, \bibinfo {author} {\bibfnamefont {L.}~\bibnamefont {Bogani}},
  \bibinfo {author} {\bibfnamefont {M.}~\bibnamefont {Bonn}}, \bibinfo {author}
  {\bibfnamefont {H.~I.}\ \bibnamefont {Wang}}, \bibinfo {author}
  {\bibfnamefont {K.}~\bibnamefont {Müllen}},\ and\ \bibinfo {author}
  {\bibfnamefont {A.}~\bibnamefont {Narita}},\ }\href
  {https://doi.org/10.1002/ange.202312610} {\bibfield  {journal} {\bibinfo
  {journal} {Angew.\ Chem.\ Int.\ Ed.}\ }\textbf {\bibinfo {volume} {62}},\
  \bibinfo {pages} {e202312610} (\bibinfo {year} {2023})}\BibitemShut {NoStop}%
\bibitem [{\citenamefont {Thongrattanasiri}\ \emph {et~al.}(2012)\citenamefont
  {Thongrattanasiri}, \citenamefont {Manjavacas},\ and\ \citenamefont
  {{Garc\'{\i}a de Abajo}}}]{thongrattanasiri2012quantum}%
  \BibitemOpen
  \bibfield  {author} {\bibinfo {author} {\bibfnamefont {S.}~\bibnamefont
  {Thongrattanasiri}}, \bibinfo {author} {\bibfnamefont {A.}~\bibnamefont
  {Manjavacas}},\ and\ \bibinfo {author} {\bibfnamefont {F.~J.}\ \bibnamefont
  {{Garc\'{\i}a de Abajo}}},\ }\href {https://doi.org/10.1021/nn204780e}
  {\bibfield  {journal} {\bibinfo  {journal} {ACS Nano}\ }\textbf {\bibinfo
  {volume} {6}},\ \bibinfo {pages} {1766} (\bibinfo {year} {2012})}\BibitemShut
  {NoStop}%
\bibitem [{\citenamefont {Cox}\ \emph {et~al.}(2016)\citenamefont {Cox},
  \citenamefont {Silveiro},\ and\ \citenamefont {{Garc\'{\i}a de
  Abajo}}}]{cox2016quantum}%
  \BibitemOpen
  \bibfield  {author} {\bibinfo {author} {\bibfnamefont {J.~D.}\ \bibnamefont
  {Cox}}, \bibinfo {author} {\bibfnamefont {I.}~\bibnamefont {Silveiro}},\ and\
  \bibinfo {author} {\bibfnamefont {F.~J.}\ \bibnamefont {{Garc\'{\i}a de
  Abajo}}},\ }\href {https://doi.org/10.1021/acsnano.5b06110} {\bibfield
  {journal} {\bibinfo  {journal} {ACS Nano}\ }\textbf {\bibinfo {volume}
  {10}},\ \bibinfo {pages} {1995} (\bibinfo {year} {2016})}\BibitemShut
  {NoStop}%
\bibitem [{\citenamefont {de~Vega}\ \emph {et~al.}(2020)\citenamefont
  {de~Vega}, \citenamefont {Cox}, \citenamefont {Sols},\ and\ \citenamefont
  {{Garc\'{\i}a de Abajo}}}]{devega2020strong}%
  \BibitemOpen
  \bibfield  {author} {\bibinfo {author} {\bibfnamefont {S.}~\bibnamefont
  {de~Vega}}, \bibinfo {author} {\bibfnamefont {J.~D.}\ \bibnamefont {Cox}},
  \bibinfo {author} {\bibfnamefont {F.}~\bibnamefont {Sols}},\ and\ \bibinfo
  {author} {\bibfnamefont {F.~J.}\ \bibnamefont {{Garc\'{\i}a de Abajo}}},\
  }\href {https://doi.org/10.1103/PhysRevResearch.2.013313} {\bibfield
  {journal} {\bibinfo  {journal} {Phys.\ Rev.\ Res.}\ }\textbf {\bibinfo
  {volume} {2}},\ \bibinfo {pages} {013313} (\bibinfo {year}
  {2020})}\BibitemShut {NoStop}%
\bibitem [{\citenamefont {{Garc\'{\i}a de
  Abajo}}(2014)}]{garciadeabajo2014graphene}%
  \BibitemOpen
  \bibfield  {author} {\bibinfo {author} {\bibfnamefont {F.~J.}\ \bibnamefont
  {{Garc\'{\i}a de Abajo}}},\ }\href {https://doi.org/10.1021/ph400147y}
  {\bibfield  {journal} {\bibinfo  {journal} {ACS Photonics}\ }\textbf
  {\bibinfo {volume} {1}},\ \bibinfo {pages} {135} (\bibinfo {year}
  {2014})}\BibitemShut {NoStop}%
\bibitem [{\citenamefont {Ashcroft}\ and\ \citenamefont
  {Mermin}(1976)}]{AM1976}%
  \BibitemOpen
  \bibfield  {author} {\bibinfo {author} {\bibfnamefont {N.~W.}\ \bibnamefont
  {Ashcroft}}\ and\ \bibinfo {author} {\bibfnamefont {N.~D.}\ \bibnamefont
  {Mermin}},\ }\href@noop {} {\emph {\bibinfo {title} {Solid State Physics}}}\
  (\bibinfo  {publisher} {Harcourt College Publishers},\ \bibinfo {address}
  {Philadelphia},\ \bibinfo {year} {1976})\BibitemShut {NoStop}%
\bibitem [{\citenamefont {Christensen}\ \emph {et~al.}(2015)\citenamefont
  {Christensen}, \citenamefont {Yan}, \citenamefont {Jauho}, \citenamefont
  {Wubs},\ and\ \citenamefont {Mortensen}}]{christensen2015kerr}%
  \BibitemOpen
  \bibfield  {author} {\bibinfo {author} {\bibfnamefont {T.}~\bibnamefont
  {Christensen}}, \bibinfo {author} {\bibfnamefont {W.}~\bibnamefont {Yan}},
  \bibinfo {author} {\bibfnamefont {A.-P.}\ \bibnamefont {Jauho}}, \bibinfo
  {author} {\bibfnamefont {M.}~\bibnamefont {Wubs}},\ and\ \bibinfo {author}
  {\bibfnamefont {N.~A.}\ \bibnamefont {Mortensen}},\ }\href
  {https://doi.org/10.1103/PhysRevB.92.121407} {\bibfield  {journal} {\bibinfo
  {journal} {Phys.\ Rev.\ B}\ }\textbf {\bibinfo {volume} {92}},\ \bibinfo
  {pages} {121407} (\bibinfo {year} {2015})}\BibitemShut {NoStop}%
\bibitem [{\citenamefont {Rasmussen}\ \emph {et~al.}(2023)\citenamefont
  {Rasmussen}, \citenamefont {{Rodr{\'\i}guez Echarri}}, \citenamefont
  {{Garc\'{\i}a de Abajo}},\ and\ \citenamefont {Cox}}]{rasmussen2023nonlocal}%
  \BibitemOpen
  \bibfield  {author} {\bibinfo {author} {\bibfnamefont {T.~P.}\ \bibnamefont
  {Rasmussen}}, \bibinfo {author} {\bibfnamefont {{\'A}.}~\bibnamefont
  {{Rodr{\'\i}guez Echarri}}}, \bibinfo {author} {\bibfnamefont {F.~J.}\
  \bibnamefont {{Garc\'{\i}a de Abajo}}},\ and\ \bibinfo {author}
  {\bibfnamefont {J.~D.}\ \bibnamefont {Cox}},\ }\href
  {https://doi.org/10.1039/D2NR06286K} {\bibfield  {journal} {\bibinfo
  {journal} {Nanoscale}\ }\textbf {\bibinfo {volume} {15}},\ \bibinfo {pages}
  {3150} (\bibinfo {year} {2023})}\BibitemShut {NoStop}%
\end{thebibliography}


\end{document}

\title{Nonreciprocal plasmons in one-dimensional carbon nanostructures
\\ {\color{gray} \small -- SUPPLEMENTARY INFORMATION --} }

\author{A.~Rodr\'{\i}guez~Echarri\,\orcidlink{0000-0003-4634-985X}}
\affiliation{Max-Born-Institut, 12489 Berlin, Germany}
\affiliation{Center for Nanophotonics, NWO Institute AMOLF, 1098 XG Amsterdam, The Netherlands}

\author{F.~Javier~Garc\'{\i}a~de~Abajo\,\orcidlink{0000-0002-4970-4565}}
\affiliation{ICFO-Institut de Ciencies Fotoniques, The Barcelona Institute of Science and Technology, 08860 Castelldefels (Barcelona), Spain}
\affiliation{ICREA-Instituci\'o Catalana de Recerca i Estudis Avan\c{c}ats, Passeig Llu\'{\i}s Companys 23, 08010 Barcelona, Spain}

\author{Joel~D.~Cox\,\orcidlink{0000-0002-5954-6038}}
\email[Corresponding author: ]{cox@mci.sdu.dk}
\affiliation{POLIMA---Center for Polariton-driven Light--Matter Interactions, University of Southern Denmark, Campusvej 55, DK-5230 Odense M, Denmark}
\affiliation{Danish Institute for Advanced Study, University of Southern Denmark, Campusvej 55, DK-5230 Odense M, Denmark}

\begin{abstract}
    We present the occupation factors of tight-binding electronic states in the graphene nanoribbons (GNRs) and carbon nanotubes (CNTs) considered in the main text when considering different charge carrier doping levels and applied in-plane static electric fields. We also provide additional simulations of GNRs interacting with with point dipole sources in other configurations, obtained using both atomistic and classical models, and investigate the interdependence of electron mobility, carrier doping levels, and drift velocity on the chemical potential and inelastic scattering rate in extended graphene.
\end{abstract}

\date{\today}
\maketitle
%\tableofcontents

\begin{figure}
    \centering
    \includegraphics[width=1\textwidth]{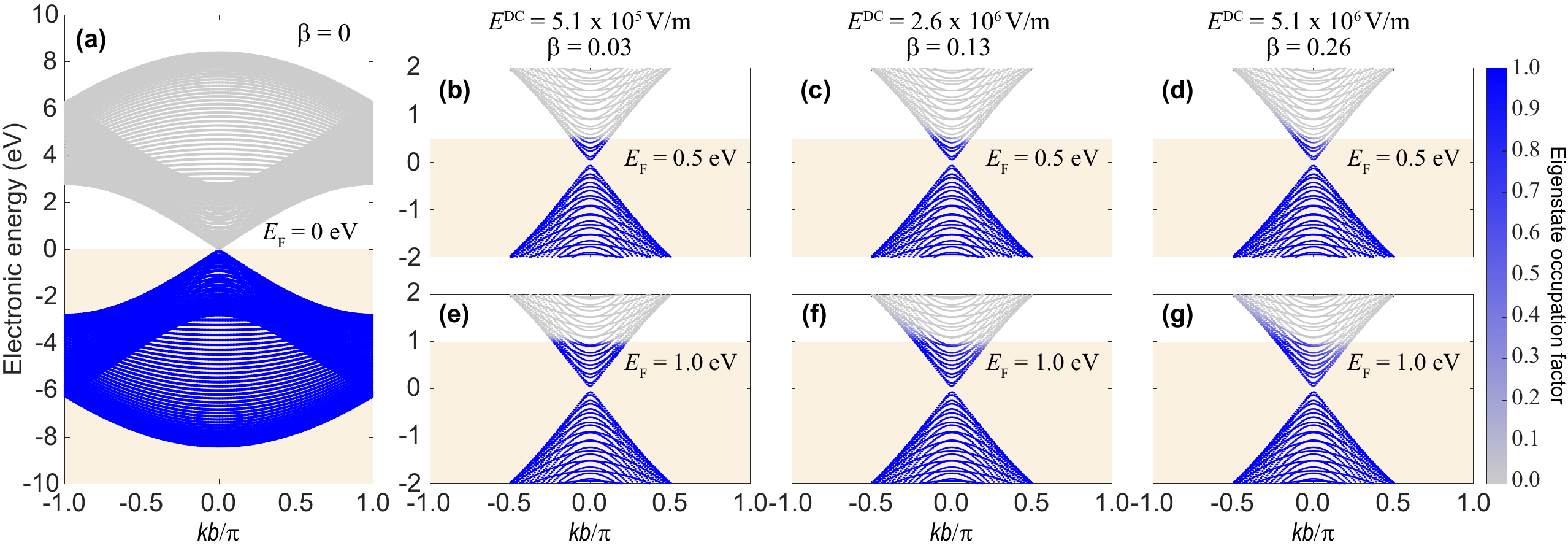}
    \caption{\textbf{Electronic bands and occupation factors in drift-biased graphene nanoribbons with armchair edge terminations.} (a) Electronic band structure of an armchair (AC) edge-terminated GNR with $10\,$nm width (curves) calculated from a nearest-neighbor tight-binding (TB) model (-2.8\,eV hopping energy) and plotted for electronic wave vectors $k$ in the first Brillouin zone. The orange shaded region indicates the filling of electronic states with energies below the Fermi energy $E_{\rm F}=0$ when the GNR is undoped and unbiased ($E^{\rm DC}=0$), while the colors of the curve indicate the occupation factors (see colorbar). Analogous plots are shown in panels (b-e) near the Dirac point when the GNR is electrically doped to $E_{\rm F}=0.5$\,eV (in the absence of applied current) and drift-biased according to the applied in-plane electric field strengths indicated above each column, along with the estimated drift velocity $v=\beta v_{\rm F}$ expressed in terms of the graphene Fermi velocity $ v_{\rm F}\approx c/300$. Panels (e-g) show the same as (b-g) but for $E_{\rm F}=1.0$\,eV. All results are obtained by assuming a mobility $\mu^{\rm DC} = 500$\,cm$^2$/(V\,s).
    }
\label{fig:figS5}
\end{figure}

\begin{figure}
    \centering
    \includegraphics[width=1\textwidth]{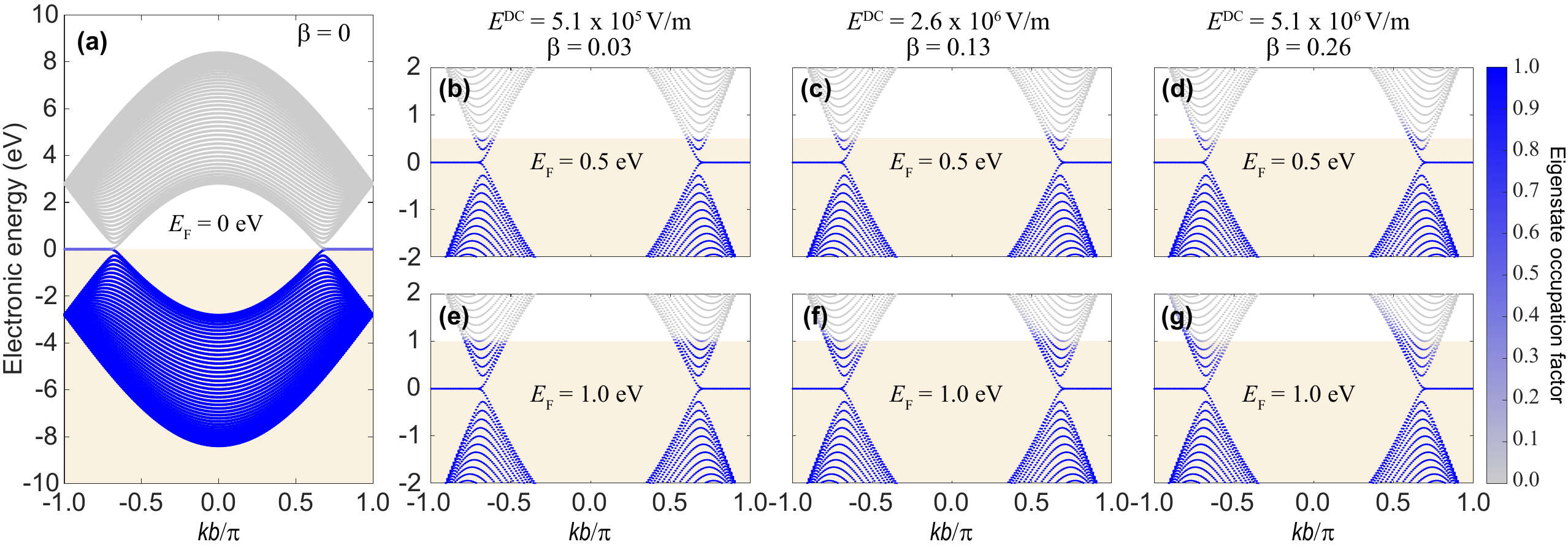}
    \caption{\textbf{Electronic states and occupation factors in drift-biased graphene nanoribbons with zigzag edge terminations.} Same as Fig.\,\ref{fig:figS5} but considering ZZ edge-terminated GNRs, which feature two Dirac cones with flat-band edge states at zero energy.
    }
\label{fig:figS6}
\end{figure}

\begin{figure}
    \centering
    \includegraphics[width=0.85\textwidth]{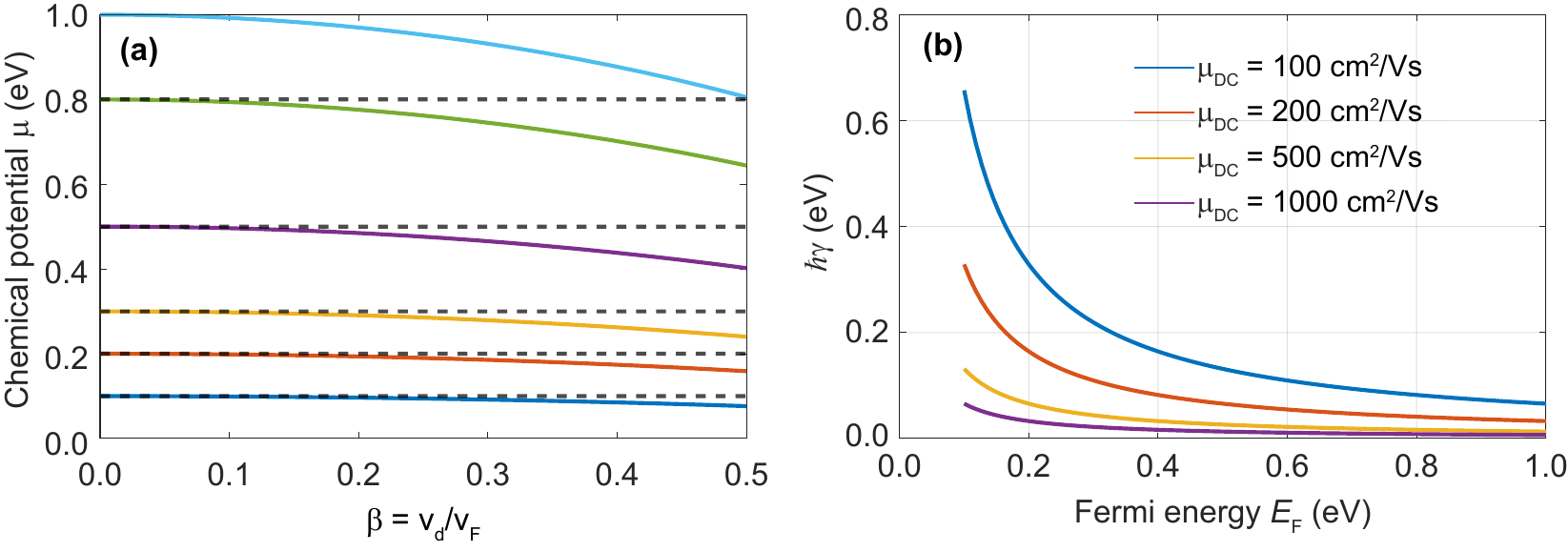}
    \caption{\textbf{Interdependence of parameters entering the conductivity of drift-biased graphene.} Here we examine the effect of drift velocity and charge carrier doping entering the semiclassical conductivity of extended graphene, as described according to the Boltzmann transport equation formalism in the main text. (a) The chemical potential is plotted as a function of the drift velocity $v = \beta  v_{\rm F}$ at the specified Fermi energies, i.e., $\mu=E_{\rm F}$ when $\beta=0$ and $T=0$, such that the initial carrier density is conserved according to the prescription in the main text (see Methods). Dashed horizontal lines indicate the considered value of $E_{\rm F}$ in each case, distinguished by a different color. (b) The inelastic scattering rate $\gamma=e  v_{\rm F}^2 /\mu_{\rm DC} E_{\rm F}$ is plotted as a function of $E_{\rm F}$ for different values of the mobility $\mu_{\rm DC}$.
    }
\label{fig:figS4}
\end{figure}

\begin{figure}
    \centering
    \includegraphics[width=0.8\textwidth]{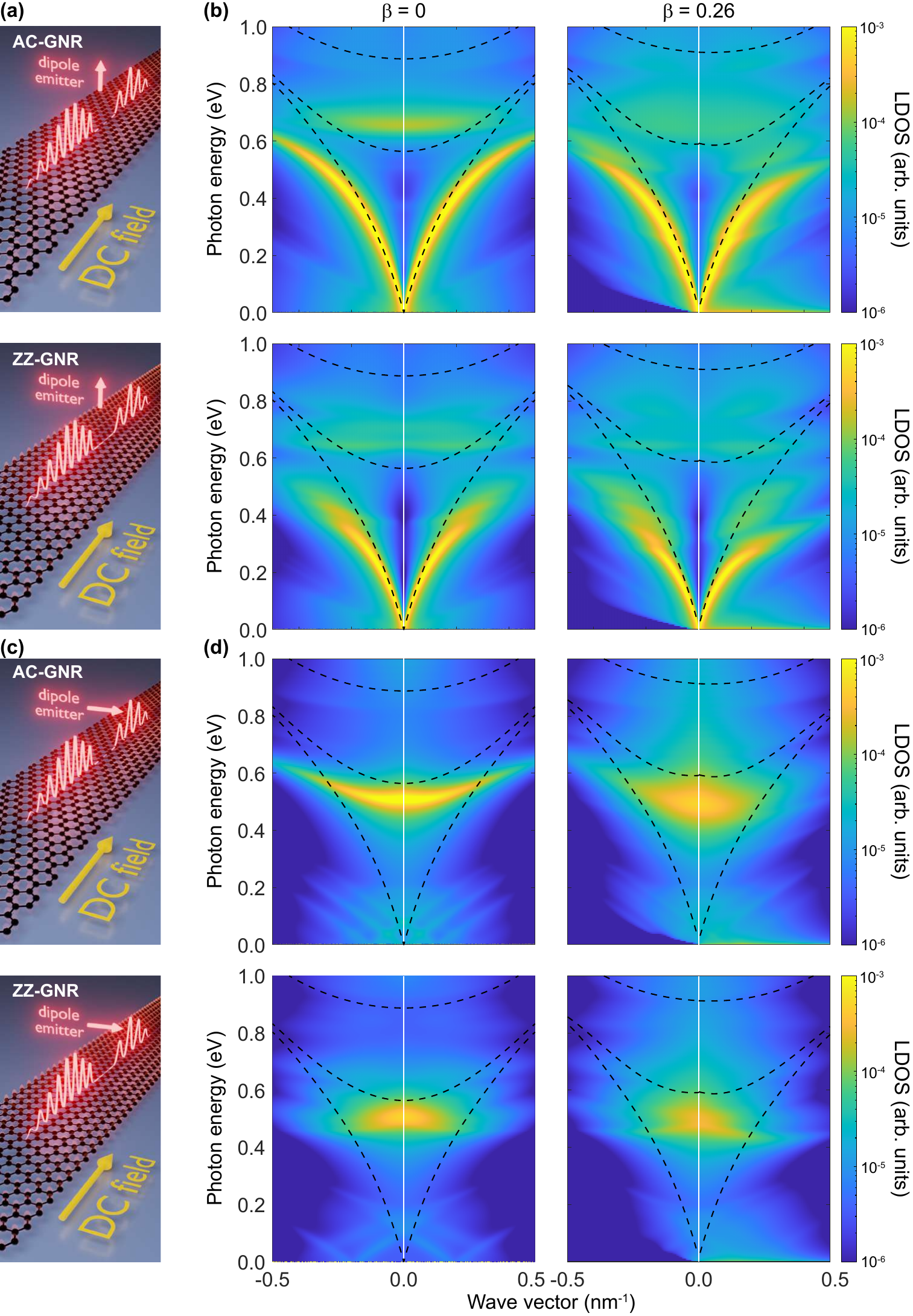}
    \caption{\textbf{Dipole coupling effects in the nonreciprocal response of drift-biased graphene nanoribbons.} (a) Schematic illustration of propagating plasmons launched in GNRs with AC (upper panel) and ZZ (lower panel) edge terminations by a dipole centered above and oriented perpendicular to the ribbon. (b) For the configuration depicted in (a), the wave-vector-resolved local photonic density of states (LDOS) is plotted as a function of photon energy when the dipole is 5\,nm above the center of $W\approx10$\,nm AC (upper row) and ZZ (lower row) GNRs when they are unbiased (left column) and biased with $E^{\rm DC}=5.2\times10^6$\,V/m (right column). Dashed curves in the contour plots indicate the plasmon dispersion predicted in the semiclassical model. Panels (c) and (d) show analogous results to (a) and (b) for a dipole oriented parallel to the GNRs. All results are obtained for $E_{\rm F}=0.5$\,eV and $\mu_{\rm DC}=500$\,cm$^2$/(V s).
    }
\label{fig:figS1}
\end{figure}

\begin{figure}
    \centering
    \includegraphics[width=1\textwidth]{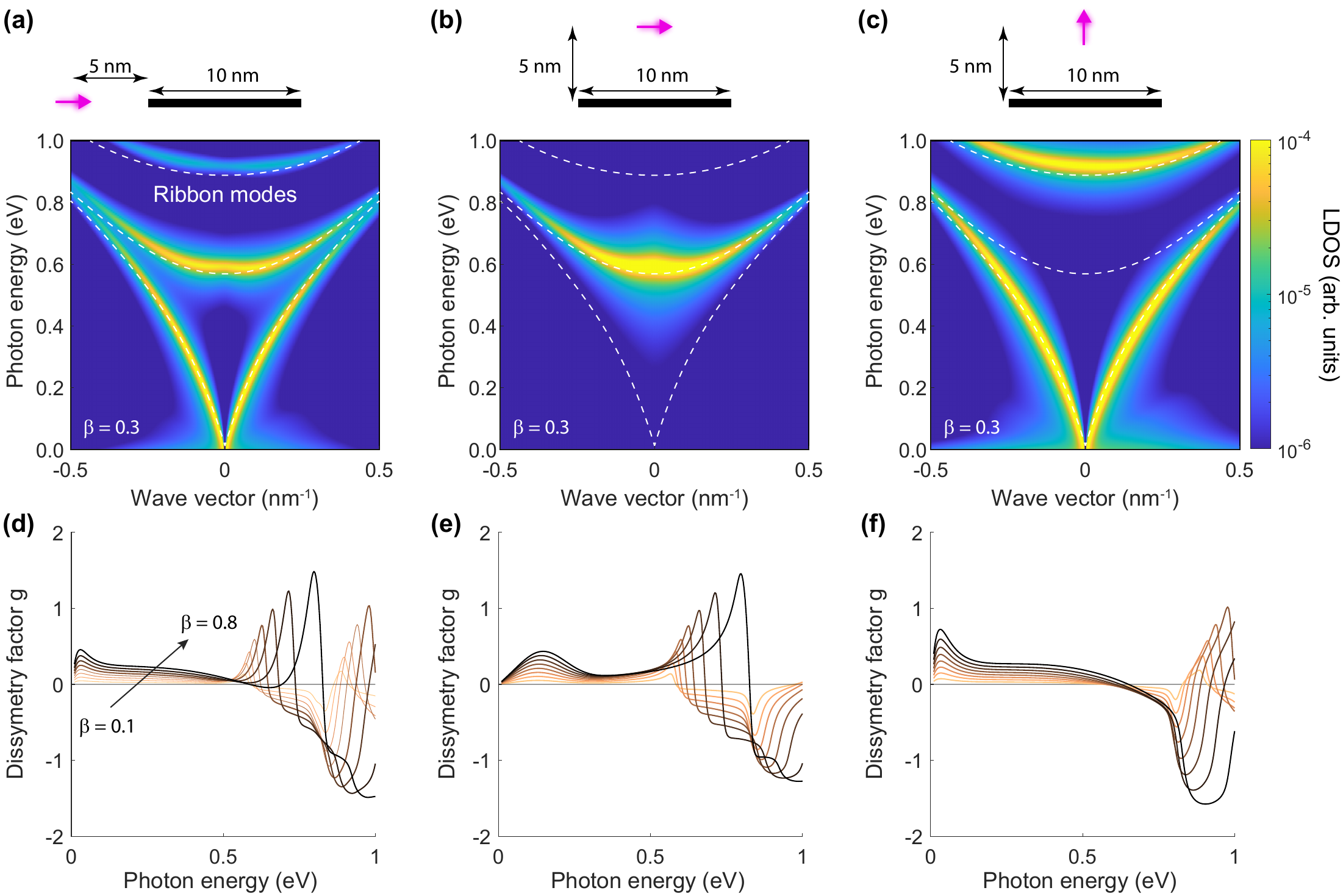}
    \caption{\textbf{Classical simulations of nonreciprocal plasmons in drift-biased graphene nanoribbons.} The local photonic density of states (LDOS), calculated using the semiclassical description of drift-biased graphene nanoribbons (GNRs) described in the Methods section of the main text, is plotted for a dipole placed 5~nm away from GNRs of 10\,nm width (a) oriented in the ribbon plane and placed to the left of its edge; (b) oriented parallel to the ribbon and located directly above its center; and (c) oriented perpendicular to the ribbon and located directly above its center (see insets above each panel). In all cases, the nanoribbons are doped with to a Fermi energy $E_{\rm F}=0.5$\,eV and the considered electron drift velocity is $\beta=v/ v_{\rm F}=0.3$, while the white dashed curves indicate the dispersion of plasmon modes in the absence of drift current. The dissymmetry factors calculated for the dipole-ribbon configurations considered in panels (a-c) are plotted in panels (d-f), respectively, as the drift velocity is varied from $\beta = 0.1$ to $\beta = 0.8$ in steps of 0.1.
    }
\label{fig:figS2}
\end{figure}

\begin{figure}
    \centering
    \includegraphics[width=0.48\textwidth]{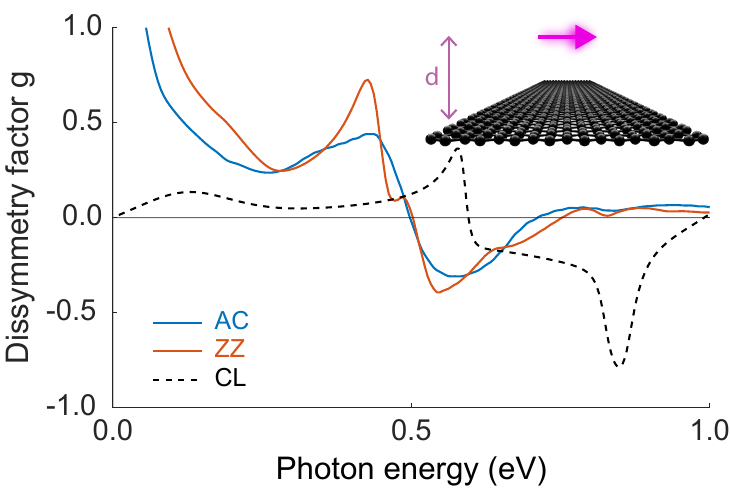}
    \caption{\textbf{Dipole coupling effects in the dissymmetry of guided plasmons in drift-biased graphene nanoribbons.} We plot the dissymmetry factors calculated through atomistic simulations (solid curves) of the armchair (AC, blue curve) and zigzag (ZZ, red curve) edge-terminated graphene nanoribbons (GNRs) with $W \approx 10$\,nm width for a dipole oriented parallel to and placed $d=5$\,nm away from the GNR plane, directly above its center (see inset). The dissymmetry factor predicted by semiclassical simulations of the same configurations is also shown (CL, black dashed curve).
    }
\label{fig:figS3}
\end{figure}

\begin{figure}
    \centering
    \includegraphics[width=0.95\textwidth]{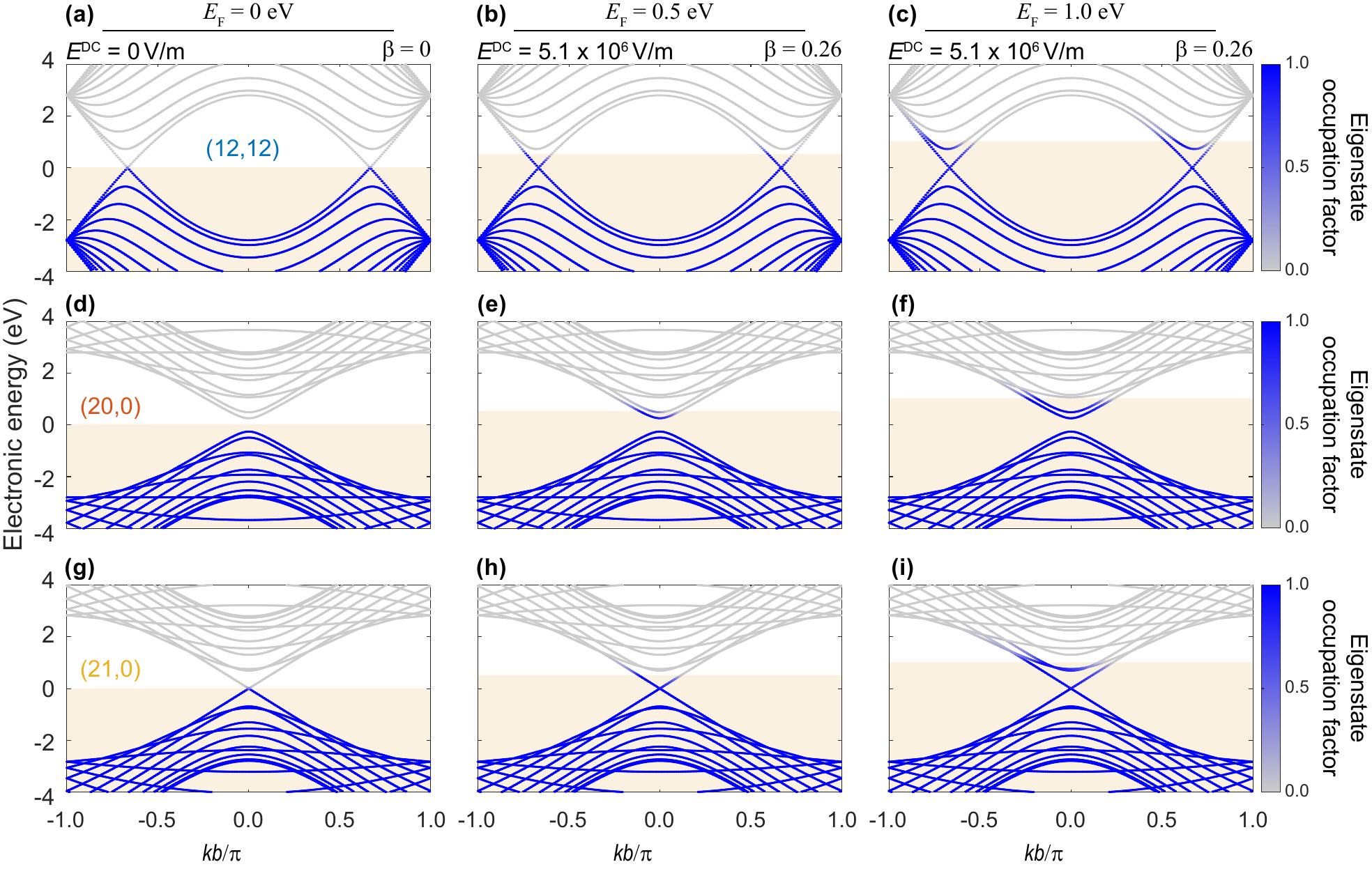}
    \caption{\textbf{Electronic bands and occupation factors in drift-biased carbon nanotubes.} The tight-binding electronic states (curves) of a (12,12) carbon nanotube (CNT) are plotted in colors corresponding to their occupation factors (see colorbar) when the CNT is (a) unbiased ($E^{\rm DC}=0$) and undoped ($E_{\rm F}=0$); (b) biased by $E^{\rm DC}=5.1\times 10^{6}$\,V/m and doped to $E_{\rm F} = 0.5$\,eV; (c) biased by $E^{\rm DC}=5.1\times 10^{6}$\,V/m and doped to $E_{\rm F} = 1.0$\,eV. The orange shaded area indicates the initial filling of electronic bands up to the Fermi energy in each considered case, while an electron mobility of $\mu_{\rm DC}=500$\,cm$^2$/(V s) is considered in all cases. In panels (d-f) and (g-i) we show analogous results to (a-c) but for the (20,0) and (21,0) CNTs, respectively.
    }
\label{fig:figS7}
\end{figure}

% \bibliographystyle{apsrev}
% \bibliography{../../../bibtex/refs} 
% \bibliography{refs}